\documentclass[%
aip,
jcp,
amsmath,amssymb,reprint,%
]{revtex4-2}

\usepackage[utf8]{inputenc}
\usepackage[T1]{fontenc}
\usepackage{microtype}

\usepackage{graphicx}
\usepackage{dcolumn}
\usepackage{multirow}
\usepackage{bm}
\usepackage{setspace}

\usepackage[breaklinks=true,colorlinks=true,linkcolor=blue,urlcolor=blue,citecolor=blue,bookmarks=true,bookmarksopenlevel=2]{hyperref}
\usepackage{xcolor}
\definecolor{navyblue}{HTML}{000080}
\usepackage[normalem]{ulem}

\usepackage{xspace}
\usepackage{textcomp}
\usepackage[textsize=footnotesize]{todonotes}
\makeatletter
\newcommand*{\Cite}[2][]{%
  \begingroup
  \let\NAT@mbox=\mbox
  \let\@cite\NAT@citenum
  \let\NAT@space\NAT@spacechar
  \let\NAT@super@kern\relax
  \renewcommand\NAT@open{}%
  \renewcommand\NAT@close{}%
  \cite[#1]{#2}%
  \endgroup
}
\makeatother

\usepackage{siunitx}

\newcommand{\ket}[1]{\left\vert #1 \right\rangle}

\newcommand{\expect}[3]{\left\langle #1 \vphantom{#2 #3} \right\vert #2 \left\vert \vphantom{#1 #2} #3 \right\rangle}
\newcommand{\propag}[2]{\langle \langle #1 \vphantom{#2}; \vphantom{#1} #2 \rangle \rangle}
\newcommand{\twoint}[2]{\langle #1 \vphantom{#2} || \vphantom{#1} #2 \rangle}

\begin{document}

\title{Relativistic and QED corrections to one-bond indirect nulcear spin-spin couplings in X$_2^{2+}$ and X$_3^{2+}$ ions (X = Zn, Cd, Hg)}
\thanks{Electronic Supplementary Information (ESI) available: basis sets used for each nuclei. See DOI: 10.1039/b000000x}

\author{Mariano Colombo Jofr\'e}%
\affiliation{Instituto de Modelado e Innovaci\'on Tecnol\'ogica (IMIT), Facultad de Ciencias Exactas, Naturales y Agrimensura, Universidad Nacional del Nordeste, Avda. Libertad 5460, W3404AAS, Corrientes, Argentina}

\author{Karol Kozio{\l}}
\affiliation{Narodowe Centrum Bada\'{n} J\k{a}drowych (NCBJ), Andrzeja So{\l}tana 7, 05-400 Otwock-\'{S}wierk, Poland}

\author{I. Agust{\'i}n Aucar}
\affiliation{Instituto de Modelado e Innovaci\'on Tecnol\'ogica (IMIT), Facultad de Ciencias Exactas, Naturales y Agrimensura, Universidad Nacional del Nordeste, Avda. Libertad 5460, W3404AAS, Corrientes, Argentina}

\author{Konstantin Gaul}
\author{Robert Berger}
\affiliation{Fachbereich Chemie, Philipps–Universit\"at Marburg, Hans-Meerwein-Stra{\ss}e 4, 35032 Marburg, Germany}

\author{Gustavo A. Aucar}
\email{gaaucar@conicet.gov.ar}
\affiliation{Instituto de Modelado e Innovaci\'on Tecnol\'ogica (IMIT), Facultad de Ciencias Exactas, Naturales y Agrimensura, Universidad Nacional del Nordeste, Avda. Libertad 5460, W3404AAS, Corrientes, Argentina}


\date{\today}

\begin{abstract}
The indirect nuclear spin-spin coupling tensor, $\bm J$, between mercury nuclei in Hg-containing systems can be of the order of few kHz and one of the largest measured. We conduct an analysis of the physics behind the electronic mechanisms that contribute to the one- and two-bond couplings $^n {\bm J}_{\mathrm{Hg}-\mathrm{Hg}}$ ($n=1, 2$).  We performed calculations for $J$-couplings in X$_2^{2+}$ and $X_3^{2+}$ ions ($X$ = Zn, Cd, Hg), within polarization propagator theory, using the random phase approximation (RPA) and the pure zeroth order approximation (PZOA), with Dirac-Hartree-Fock (DHF) and Dirac-Kohn-Sham (DKS) orbitals, both at four-component and ZORA levels. We show that the ``paramagnetic-like'' mechanism contribute with more than 99.98\% to the total isotropic component of the coupling tensor. By means of an analysis of the molecular and atomic orbitals involved in the total value of the response function, we find that the $s$-type valence atomic orbitals have a predominant role in the description of the coupling. This fact allows us to develop an effective model from which quantum electrodynamics (QED) effects on $J$-coupling in the aforementioned ions can be estimated. The estimated QED corrections were found in the interval $(0.7; ~ 1.7)$\% of the total relativistic effect on isotropic one-bond $^1 {\bm J}$ coupling and from the interval $(-0.2; ~ -0.4)$\%, in Zn-containing ions, to $(-0.8; ~ -1.2)$\%, in Hg-containing ions, of the total isotropic coupling constant in the studied systems. We also show that estimated QED corrections cast a visible dependence on the nuclear charge $Z$ of each atom $X$ in the form of a power-law $\propto Z^5$. 

\end{abstract}

\maketitle



\section{Introduction}
\label{Introduction}

Nuclear magnetic resonance (NMR) spectroscopy is a powerful experimental technique used, among many other applications, to identify chemical compounds and predict their molecular structures.
Precise calculations of two of its most relevant spectroscopic parameters, the NMR shielding constant, $\sigma$, and the indirect nuclear spin-spin coupling  constant, $J$, are highly challenging. They require to consider several intramolecular effects (and in condensed phases also intermolecular effects) with the proper theories and state-of-the-art models.\cite{Kaupp2004, JVaara_PCCP2007, Contreras2013, LKrivdin_RAS2013}

There are few leading electronic effects, such as electron correlation and relativistic effects, that should be included in order to get an accurate theoretical reproduction of the nuclear magnetic shieldings. In the case of heavy atom-containing molecules it is known that relativistic effects may be as large as the non-relativistic (NR) contributions.\cite{Visscher1999,GAA_IRPC2010}

Among the different formalisms that were developed to introduce QED effects on atomic systems,\cite{VShabaev_PhysRep2002, ILindgren_PhysRep2004, KGDyall_JCP2013, ILindgren_Chap9-Handbook2017, VShabaev_CompPhysComm2018} there is the polarization propagator which was recently derived from the path integral version of quantum theory.\cite{GAA_PCCP2014} This fact gives new insights on how to include QED and correlation effects altogether, through the consideration of the effects of external perturbations on a many-body quantum system that is described within a QED-based theoretical framework. As sketched in Ref. \citenum{GAA_IJQC2019}, once the appropiate generating functional is defined one can derive from it the renormalized propagators that include QED effects together with electron correlation. One should be aware of the fact that the fluctuation potential does introduce the internal interactions in the generating functional. The external perturbations that shall be considered for the calculation of any given response properties are related to the perturbative potential that shall be taken together with basic excitations, being the principal propagator related with the functional derivatives of the connected generating functional with respect to those external perturbations. 

Recently we published preliminary results concerning the estimation of QED effects on NMR shielding constants for He-like and Be-like atomic systems with $10 \leq Z \leq 86$.\cite{Gimenez2016} In that work we presented a model in which QED corrections, obtained by Yerokhin {\it et al.}\cite{Yerokhin2011,Yerokhin2012} for H-like atoms, are scaled to the aforementioned ionic systems. Such procedure is similar to the way QED effects are usually introduced in multi-electron atoms.\cite{Lowe2013} As a next step in our research program, we estimated QED effects on shielding of neutral and ionic atoms with $10 \leq Z \leq 86$, and diatomic halogen molecules using an extension of our previous approach.\cite{Karol_JCP2019} 
To our knowledge there is no other estimation of QED effects on nuclear shielding of molecular systems. Our results show that QED effects are significant and should be taken into account in order to obtain calculated values closer to experiment. At the moment, highly accurate absolute values of NMR shieldings in some gas-phase molecules can be obtained by experiments. Their error bars may be less than the values of QED corrections for heavy-atom containing molecules.\cite{KJackowski_JMolStruct2005, KJackowski_JPCA2010, KJackowski_Prog2012, KJackowski_PCCP2016}

In the present work, we extend the previous approach to estimate the order of magnitude of QED effects on indirect nuclear spin-spin couplings, $\bm J$. This property is better suited for the analysis of QED effects because its experimental measurement is simpler than that of NMR shieldings, making the contrast between theoretical predictions and experiments more straightforward. Moreover, $\bm J$ tensors can be linked to the hyperfine structure (hfs) of atoms, for which QED corrections have been published for few-electron systems in some papers, for H-like ions in e.g. Refs. \citenum{Sapirstein_PRA1997, Soff_PRA1998, Karshenboim_PLB2002} and for Li-like ions in e.g. Refs. \citenum{Indelicato_EurPhysJD2000, Sapirstein_PRA2001}. For neutral atoms there are only few works focusing on QED effects to hfs. In the case of alkali-metal atoms, available data are for $s$ states,\cite{Sapirstein_PRA2003,Ginges2017} $p_{1/2}$ states\cite{Sapirstein_PRA2006} and $p_{3/2}$ states.\cite{Sapirstein_PRA2008}
Pyykk{\"o} and Zhao \cite{Pyykko2003} suggested, basing their study on the self-energy (SE) effects on magnetic dipole hyperfine integrals, that the SE effect should be of the order of -3\% for the $J$(Hg-Hg) coupling. 
The vacuum polarization (VP) effect to hfs in the atoms and ions having valence $s$ orbitals is about a half of the SE effect for $Z\approx80$ and have opposite signs.\cite{Sapirstein_PRA2003,Ginges2017} Then, one can conclude that the total QED effect on the $J$(Hg-Hg) coupling should be of the order of -1\%. 

One of the aims of this work is to apply an effective model to estimate QED effects to $\bm J$ tensors in the model ions $X_2^{2+}$ and $X_3^{2+}$ ($X$ = Zn, Cd, Hg). This choice was made on the fact that the $J$-coupling between mercury nuclei in Hg-containing systems can reach an order of few kHz.\cite{Autschbach_JACS2003,Autschbach_JACS2007} This comparatively sizeable amount makes them good candidates to learn about one of the smallest effects that may be observed, giving new understandings on the physics beyond the usual relativistic quantum chemistry. We also show the $Z$-dependent patterns that the relativistic and QED effects follow in all these systems.

Another aim of this work is to analyze the influence of relativistic effects on the just mentioned $J$-couplings by comparison of calculations on the level of two-component \textit{zeroth order regular approximation} (ZORA). We also investigate the effect of electron correlation by using \textit{density functional theory} (DFT) in different flavors from \textit{local density approximation} (LDA), over \textit{generalized gradient approximation} (GGA) functionals to hybrid DFT.
%
\section{Theoretical models and computational details}
%

%
\subsection{Spin-spin coupling and polarization propagator theory}
%
Within the four-component polarization propagator formalism,\cite{GAA_IRPC2010} the indirect nuclear spin-spin coupling between nuclei $K$ and $L$ is a tensor ${\bm J}(K, L)$ whose components may be written as
follows (in SI units)\cite{GO93}
\begin{equation}
\label{eq:JKL}
{\bm J}_{\mu \nu}(K, L) = \left( \frac{\mu_0}{4\pi} \right)^2 \frac{e^2 \hbar^2}{h} \gamma_K \gamma_L \propag{V_{\mu}^K}{V_{\nu}^L}
\end{equation}
where $\mu, \nu = x, y, z$, $\mu_0$ is the magnetic constant, $e$ is the elementary electric charge, $\hbar=h/(2\pi)$ is the reduced Planck constant $h$, $\gamma_K$ and $\gamma_L$ are the magnetogyric ratios of nuclei $K$ and $L$. ${\bm J}_{\mu \nu}(K, L)$ is written in terms of the response function,\cite{GO93} $\propag{V_{\mu}^K}{V_{\nu}^L}$, defined for the {\it hyperfine structure} (hfs) operators for nuclei $K, L$, whose components are
\begin{equation}
\label{eq:V}
V_{\mu}^X = \left( \frac{{\bm r}_X \times c{\bm \alpha}}{r_X^3} \right)_{\mu}
\end{equation}
where ${\bm \alpha} = \left(\alpha_x, \alpha_y, \alpha_z\right)$ are the 4$\times$4 Dirac matrices, which are written in the standard representation in terms of the 2$\times$2 Pauli matrices ${\bm \sigma} = \left(\sigma_x, \sigma_y, \sigma_z\right)$ as
\begin{equation}
{\bm \alpha} =
\begin{pmatrix}
{\bm 0} & {\bm \sigma} \\
{\bm \sigma} & {\bm 0}
\end{pmatrix}
\end{equation}
and ${\bm r}_X = {\bm r} - {\bm R}_X$, the relative electron position with respect to nucleus $X$ ($X$ = K, L). Operator \eqref{eq:V} is related to the magnetic interaction between an electron belonging to the molecular electronic framework and the nuclear magnetic dipole moment, causing hyperfine splitting of the molecular energy levels.

Each component ${\bm J}_{\mu \nu}(K, L)$ may also be written in terms of two different kind of objects\cite{GAA_IRPC2010} known as the {\it perturbators}, {\bf b}, and the {\it principal propagator}, {\bf P}, such that
\begin{equation}
{\bm J}_{\mu \nu}(M,N) = \frac{1}{2} \left( {\bm b}_{\mu}^M ~ {\bm P} ~ {\bm b}_{\nu}^N + c.c. \right)
\end{equation}
where the perturbator matrix elements have the following formal expression
\begin{equation}
\label{eq:perturbators}
\left( {\bm b}_{\mu}^X \right)_{ia} = \left(\frac{\mu_0}{4\pi}\right)
\frac{e\hbar}{\sqrt{h}} \gamma_X \expect{i}{\left( \frac{{\bm r}_X
\times c{\bm \alpha}}{r_X^3} \right)_{\mu}}{a}
\end{equation}
and are related with the transition moments between occupied
$i$ and unoccupied $a$ molecular orbitals with the hfs operator written in a molecular orbital basis. This representation can also be expressed in terms of atomic orbitals in such a way that eq. \eqref{eq:perturbators} can be written as a linear combination of an atomic orbital-based representation of the hfs operator.

At first consistent order of approach, also known as {\it random phase approximation} (RPA), the principal propagator matrix is given by\cite{GAA_IRPC2010}
\begin{equation}
{\bf P} =
 \begin{pmatrix}
 \boldsymbol A & {\boldsymbol B}^* \\
 \boldsymbol B & {\boldsymbol A}^*
\end{pmatrix}^{-1}
\end{equation}
where the elements of matrices {\bf A} and {\bf B} are written, within the second-quantization formalism, as
\begin{align}
\label{eq:A01}
\boldsymbol A_{ia,jb} & = -\expect{0}{[a_i^\dagger a_a,[a_b^\dagger a_j, H_0]]}{0} \nonumber \\
& = \delta_{ab}\delta_{ij}(\varepsilon_a - \varepsilon_i) + \tilde{G}_{ajib} \nonumber  \\
& = A(0)_{ia,jb} + A(1)_{ia,jb}
\end{align}
and
\begin{align}
\label{eq:B1}
\boldsymbol B_{ia,jb} & = - \expect{0}{[a_a^\dagger a_i,[a_b^\dagger a_j, H_0]]}{0} = - \tilde{G}_{jiab} \nonumber \\
& = B(1)_{ia,jb}\,,
\end{align}
where $\tilde{G}_{ajib}$ is the two-electron tensor in molecular orbital basis.  In the general case of hybrid DFT, $\tilde{G}_{ajib}$ is constructed as
\begin{equation}
\tilde{G}_{ajib}=\langle aj|ib\rangle-a_{\mathrm{X}}\langle aj|bi\rangle
+a_\mathrm{DFT}\langle aj|\hat{V}'_\mathrm{XC}[\rho,\bm{\nabla} \rho]|ib\rangle\,.
\label{eq: gmatrix_aobasis}
\end{equation}
Here $\hat{V}'_\mathrm{XC}$ is the functional derivative of the exchange correlation potential operator with respect to the electronic density function $\rho = \sum_i\phi_i^\dagger\phi_i$, where the sum is over all occupied spinors, and for GGA functionals with respect to the gradient of the density $\bm{\nabla} \rho$. The total density function $\rho$ can be separated in spin-free and spin-dependent parts. In case of pure DFT (non-hybrid) we have $a_\mathrm{X}=0$ and in case of pure HF we have $a_\mathrm{X}=1$ and $a_\mathrm{DFT}=0$. In the latter case
$\tilde{G}_{ajib}$ reduces to $\twoint{aj}{ib}=\langle aj|ib\rangle - \langle aj|bi\rangle$, with $\langle aj|ib\rangle=\langle\phi_a\phi_j|1/r_{12}|\phi_i\phi_b\rangle$. Subscripts $a,b,\ldots$ refer to unoccupied molecular orbitals, whereas $i,j,\ldots$ stands for occupied molecular orbitals. $H_0$ refers to the unperturbed electronic Hamiltonian.

Within the relativistic regime the set of unoccupied orbitals is split into two subsets, which span the positive and negative branch of energies. Excitations from occupied electronic states to negative ($ep$) and positive ($ee$) energy solutions are related with the diamagnetic-like, $J^{pp}$, and paramagnetic-like, $J^{ee}$, contributions.\cite{GAA99, GAA_PCCP2014}

As mentioned in Ref. \citenum{Gimenez2016}, actual calculations are not performed using eq. \eqref{eq:JKL} but an algorithm is used, which solves the product among the principal propagator (the inverse of the electronic Hessian) and the right perturbator column matrix, in a self-consistent manner. Therefore, the actual calculation of the coupling is performed according to the following equation
\begin{align}
{\bm J}_{\mu \nu}(K, L) & = \sum_{ia, jb}
\begin{pmatrix}
\left( {\bm b}_{\mu}^K \right)_{ia} & \left( {\bm b}_{\mu}^K \right)_{ia}^*
\end{pmatrix}
\left({\bm P}\right)_{ia, jb} 
\begin{pmatrix}
\left( {\bm b}_{\nu}^L \right)_{jb}^*  \\
\left( {\bm b}_{\nu}^L \right)_{jb}
\end{pmatrix} \nonumber \\
\label{eq:bPb-reduced}
& = \sum_{ia}
\begin{pmatrix}
\left( {\bm b}_{\mu}^K \right)_{ia} & \left( {\bm b}_{\mu}^K \right)_{ia}^*
\end{pmatrix}
\begin{pmatrix}
\left( {\bm X}_{\nu}^L \right)_{ia}^*  \\
\left( {\bm X}_{\nu}^L \right)_{ia}
\end{pmatrix} \nonumber \\
& = \sum_{ia} [{\bm J}_{\mu \nu}(K, L)]_{ia}
\end{align}
All the information related to the principal propagator and one of the two perturbators is contained in the  ${\bf X}$ column matrix
\begin{equation}
\begin{pmatrix}
\left( {\bm X}_{\nu}^L \right)_{ia}^*  \\
\left( {\bm X}_{\nu}^L \right)_{ia}
\end{pmatrix} =
\sum_{jb} \left({\bm P}\right)_{ia, jb} 
\begin{pmatrix}
\left( {\bm b}_{\nu}^L \right)_{jb}^*  \\
\left( {\bm b}_{\nu}^L \right)_{jb}
\end{pmatrix}
\end{equation}
The way it is formally derived and implemented in the four-component DIRAC code is explicitly given in ref. \citenum{Saue-linear-response}. Each individual term $[{\bm J}_{\mu \nu}(M,N)]_{ia}$ determines a particular {\it one-particle coupling pathway} because they are defined by means of two molecular orbitals instead of four (two occupied and two unoccupied) as is usually done in our formalism.\cite{GAA_IRPC2010}
%
\subsection{Relativistic and electron correlation effects}
%
In the first paper about relativistic polarization propagators one of the main advantages of the formalism was shown, namely the possibility to get the value of relativistic effects by setting the speed of light $c \to \infty$ in the model itself.\cite{GAA_IJQC1993} Another advantage is that relativistic effects can be analyzed separately on each of the terms appearing in eq. \eqref{eq:bPb-reduced} with the same procedure. Then, by considering perturbators and principal propagator as separated objects, it is possible to obtain different information about the origin of the perturbation that takes place in the regions close to the coupled nuclei together with the efficiency of the transmission of those perturbations through the molecular electronic framework. In addition, this scheme facilitates the elucidation of what kind of relativistic effects (e.g. scalar or spin-dependent) mostly affect the matrix elements involved in eq. \eqref{eq:bPb-reduced}.


Another aspect polarization propagator theory makes possible is the account for electron correlation to different orders in the calculation of response properties. This gets clearer from eqs. \eqref{eq:A01} and \eqref{eq:B1}. Keeping the diagonal matrix {\bf A}$(0)$ alone, and neglecting matrices {\bf A}$(1)$ and {\bf B}$(1)$ is equivalent to have a one-particle operator Hamiltonian $H_0$, which is diagonal in the considered spin-orbital basis (i.e. $H_0$ describes a system of non-interacting particles). This is known as the
\textit{pure zeroth-order approximation} or PZOA. Matrices {\bf A}$(1)$ and {\bf B}$(1)$ arise when two-particle interactions are considered as part of $H_0$ and are, therefore, the resulting expressions consistent through first order in the electron interaction.
Electron correlation, when viewed in this picture, thus, manifests itself in the occurrence of two-electron integrals in eqs. \eqref{eq:A01} and \eqref{eq:B1}. This is a result of the propagator formalism and arises whether the set of spin-orbitals used in the construction of the reference state wave function, $\ket{0}$, is obtained by solving the Hartree--Fock or the Kohn--Sham equations self-consistently.
%
\subsection{Estimating QED effects on spin-spin coupling constants}\label{sec:model}
%
The approach followed in this work lies on similar grounds to that applied in our previous papers,\cite{Gimenez2016,Karol_JCP2019} in which a model to estimate QED corrections on the NMR shieldings of some light and heavy atoms and ions, as well as, some homo-nuclear diatomic molecules, was proposed. In the latter, we considered that leading QED corrections to both, perturbators and principal propagator, are enough to estimate an order of magnitude for QED corrections on spin-spin couplings.\cite{GAA_PCCP2014}

The model we developed, and is explained in detail below, allows a reliable appreciation of the magnitude of QED effects on one-bond indirect nuclear spin-spin coupling $^1 J$ in homonuclear ionic systems of the type X$_2^{2+}$ and X$_3^{2+}$, with elements X = Zn, Cd, Hg of group 12 in the periodic table since the following characteristics are found on these systems: a) there are one or two main coupling pathways that contribute with more than 70\%
of the total value of the spin-spin coupling; b) the corrections due to QED effects on the elements of each of the perturbators centered on semi-heavy or heavy nuclei represent a far greater contribution to the total value of the coupling than that coming from corrections on the matrix elements of the principal propagator; and c) the individual elements of each perturbator are dominated by $s$-type atomic hfs matrix elements.

All ions considered in this work proved to fulfill item a) for the diagonal components $^1 {\bm J}_{\mu \mu}(K, L)$ of the coupling tensor under all circumstances, being the preponderant coupling pathway that involving HOMO and LUMO. Just one exception to this statement was found in the set of calculations corresponding to one-bond coupling in X$_3^{2+}$ ions, where the main contribution represents about 50\% of the total coupling (see Table \ref{tab:total-values}). Therefore, the subsequent derivations will be concerned on the estimation of QED corrections on this particular coupling path associated to the components $^1 {\bm J}_{\mu \mu}(K, L), ~ \mu = x, y, z$ in X$_2^{2+}$ ions. Little modifications (which will be explained at the beginning of section \ref{sec:results}) must be made for the case of X$_3^{2+}$ ions.

First, we recall that at zeroth order (PZOA) we have
\begin{align}
& {\bm J}_{\mu \mu}(K, L) = 2 \Re \left( \sum_{ia} \frac{\left( {\bm b}_{\mu}^K \right)_{ia} \left( {\bm b}_{\mu}^L \right)_{ia}^*}{\epsilon_a - \epsilon_i} \right) \nonumber \\
\label{eq:J-PZOA}
& = \left(m_{hl}\right)_{\mu \mu} + 2 \Re \left(
\sum {'} \frac{\left( {\bm b}_{\mu}^K \right)_{ia} \left( {\bm b}_{\mu}^L \right)_{ia}^*}{\epsilon_a - \epsilon_i} \right)
\end{align}
where in the second line the sum $\sum {'}$ runs over all $a,i$ without that $a=l$ and $i=h$ is fulfilled at the same time. Here we have separated from the general sum the HOMO-LUMO amplitude
\begin{equation}
\label{eq:HOMO-LUMO-PZOA}
\left(m_{hl}\right)_{\mu \mu} = 2 \Re {\frac{\expect{\phi_l}{V_{\mu}^K}{\phi_h}\expect{\phi_h}{V_{\mu}^L}{\phi_l}}{{\epsilon_l - \epsilon_h}}}.
\end{equation}
Then we consider that HOMO and LUMO wave functions can be written as linear combinations of atomic orbitals
\begin{widetext}

\begin{eqnarray}
\label{eq:HOMO}
\ket{\phi_h} & = \Big( \sum_A c_{ns,h}^{A} \ket{\chi_{ns}^A} + c_{\bar{ns},h}^{A} \ket{\chi_{\bar{ns}}^A} \Big) + \sum_A \sum_{p \neq ns}c_{p,h}^A \ket{\chi_p^A} + c_{\bar{p},h}^A \ket{\chi_{\bar{p}}^A} + \ket{\psi_h^{pol}} \\
\label{eq:LUMO}
\ket{\phi_l} & = \Big( \sum_A c_{ns,l}^{A} \ket{\chi_{ns}^A} + c_{\bar{ns},l}^{A} \ket{\chi_{\bar{ns}}^A} \Big) + \sum_A \sum_{p \neq ns}c_{p,l}^A \ket{\chi_p^A} + c_{\bar{p},l}^A \ket{\chi_{\bar{p}}^A} + \ket{\psi_l^{pol}}
\end{eqnarray}

\end{widetext}
where $n$ stands for the valence principal quantum number of each element ($n = 4$ for zinc, $n = 5$ for cadmium and $n = 6$ for mercury) and a bar over the quantum numbers indicates the Kramers partner. Wave functions $\chi_p^A, \chi_{\bar{p}}^A$ are also known as reference orbitals at center $A$ and $\psi_h^{pol}, \psi_l^{pol}$ represent the part of each molecular wave function that is not spanned by the set of reference orbitals and is orthogonal to the sums in equations \eqref{eq:HOMO} and \eqref{eq:LUMO}.\cite{Faegri_JCP2001, Dubillard_JCP2006}

Under the assumption of a small contribution from the polarization terms and all other atomic orbitals apart from $ns$, we further approximate equations \eqref{eq:HOMO} and \eqref{eq:LUMO} leaving just the terms between parenthesis in the above expansions; each of these consists in four terms corresponding to both coupled nuclei
\begin{align}
\label{eq:H-aprox}
\ket{\phi_h} & \approx c_{ns,h}^K \ket{\chi_{ns}^K} + c_{\bar{ns},h}^K \ket{\chi_{\bar{ns}}^K} + c_{ns,h}^L \ket{\chi_{ns}^L} + c_{\bar{ns},h}^L \ket{\chi_{\bar{ns}}^L} \\
\label{eq:L-aprox}
\ket{\phi_l} & \approx d_{ns,l}^K \ket{\chi_{ns}^K} + d_{\bar{ns},l}^K \ket{\chi_{\bar{ns}}^K} + d_{ns,l}^L \ket{\chi_{ns}^L} + d_{\bar{ns},l}^L \ket{\chi_{\bar{ns}}^L}
\end{align}
Expansion coefficients in eqs. \eqref{eq:H-aprox} and \eqref{eq:L-aprox} were obtained with the use of \textit{projection analysis}\cite{Dubillard_JCP2006, Saue_JCP2020} combined with the \textit{intrinsic atomic orbital (IAO)} approach\cite{Knizia_JCTC2013} (see more details about how the projection analysis was performed in section \ref{sec:comp-det}). The validity of the above approximations is confirmed by the predominant ``$s$ atomic-character'' of the molecular orbitals involved, which is evidenced in the results of the projection analysis (all expansion coefficients used in this work are provided in the Supplementary Information).

From eqs. \eqref{eq:H-aprox} and \eqref{eq:L-aprox} the perturbator matrix elements can be estimated, taking into account that: i) matrix elements involving distinct wave functions (either with different centers or associated to different Kramers partners, like as $\expect{\chi_{ns}^K}{V^K}{\chi_{ns}^L}$ or $\expect{\chi_{ns}^K}{V^K}{\chi_{\bar{ns}}^L}$) are very small or vanish; ii) expectation values constructed with wave functions and operators associated with different centers, like $\expect{\chi_{ns}^L}{V^K}{\chi_{ns}^L}$, are very small or vanish; iii) expectation values of hfs operators with respect to Kramers partners centered at the same nucleus are equal (for example, $\expect{\chi_{ns}^K}{V^K}{\chi_{ns}^K} = \expect{\chi_{\bar{ns}}^K}{V^K}{\chi_{\bar{ns}}^K}$). Under these considerations, we arrive at the final approximate form of HOMO-LUMO perturbator matrix elements
\begin{align}
\label{eq:pert-aprox-K}
\boldsymbol{b}_{hl}^K & \approx \left[ (d_{ns,l}^K)^* {c_{ns,h}^K} + (d_{\bar{ns},l}^K)^* {c_{\bar{ns},h}^K} \right] \langle V \rangle_{ns} \\
\label{eq:pert-aprox-L}
{\boldsymbol{b}_{hl}^L}^{*} & \approx \left[ (c_{ns,h}^L)^* {d_{ns,l}^L} + (c_{\bar{ns},h}^L)^* {d_{\bar{ns},l}^L} \right] \langle V \rangle_{ns}
\end{align}
where $\langle V \rangle_{ns} = \expect{\chi_{ns}^X}{V^X}{\chi_{ns}^X} = \expect{\chi_{\bar{ns}}^X}{V^X}{\chi_{\bar{ns}}^X}$, $X = K, L$ corresponds to the radial integral (in atomic units $a_0^{-2}$)
\begin{equation}
\label{eq:int-r}
R^{(-2)}\left(n_1 \kappa_1 n_2 \kappa_2\right) = \int_0^{\infty} r^{-2} \left(P_{n_1 \kappa_1} Q_{n_2 \kappa_2} +Q_{n_1 \kappa_1} P_{n_2 \kappa_2} \right) dr
\end{equation}
being $P_{n \kappa}, ~ Q_{n \kappa}$ the radial parts of the one-electron wave function as defined in eqs. (11), (13) and (14) of Ref. \citenum{Gimenez2016} and we ommited subscript $\mu$ because of the spherical symmetry of the atomic wave functions $\chi_{ns}^X, \chi_{\bar{ns}}^X$. The absolute values of the $\langle V \rangle_{ns}$
integrals were obtained from MCDFGME code,\cite{mcdfgme-web} analogously to what had been done in a previous work.\cite{Karol_JCP2019} In the present case, we sought for hfs integrals associated to nucleus X with corresponding
$ns$ valence orbital. All values from MCDFGME correspond to Dirac-Coulomb-Breit Hamiltonian without including Uehling potential and were multiplied by the inverse of the fine structure constant (in atomic units), $c_0 = (1/\alpha) a_0 E_\mathrm{h}/\hbar$, in order to make the amplitudes comparable to those reported by DIRAC. The values of $\langle V \rangle_{ns}$ integrals used in the present work are listed in the Supplementary Information.

Gathering together eqs. \eqref{eq:J-PZOA}, \eqref{eq:pert-aprox-K} and \eqref{eq:pert-aprox-L} we arrive at the following expression for the approximate diagonal HOMO-LUMO coupling amplitudes at PZOA (in atomic units $a_0^{-2} \left(\hbar/E_h\right)^{-2} E_h^{-1}$)
\begin{equation}
\label{eq:HL-amp-aprox}
\left(m_{hl}\right)_{\mu \mu} \approx 2 \gamma \frac{\left(c_0 \langle V \rangle_{ns}\right)^2}{\epsilon_l - \epsilon_h}
\end{equation}
where expansion coefficients are collected in the constant
\begin{align}
\gamma = \Re \Big\{ & \left[ (d_{ns,l}^K)^* {c_{ns,h}^K} + (d_{\bar{ns},l}^K)^* {c_{\bar{ns},h}^K} \right] \times \nonumber \\
\label{eq:gamma}
& \left[ (c_{ns,h}^L)^* {d_{ns,l}^L} + (c_{\bar{ns},h}^L)^* {d_{\bar{ns},l}^L} \right]\Big\}
\end{align}
The performance of this approximation is shown in Table \ref{tab:HL-amp-aprox}, where we compare the results given by DIRAC with the value given by eq. \eqref{eq:HL-amp-aprox}. An issue related to the projection analysis impeded the estimation of QED corrections for BH\&HLYP results in the case of Hg$_2^{2+}$.
\begin{table}[ht!]
\setlength{\extrarowheight}{1.5mm}
\caption{Comparison between HOMO-LUMO contributions to the isotropic one-bond coupling $^1 J_\mathrm{iso}^\mathrm{ee}$ as given by DIRAC, $^1 \left(m_{hl}\right)_\mathrm{iso}$, and approximate values given by eq. \eqref{eq:HL-amp-aprox}. The amplitudes associated to X$_3^{2+}$ ions correspond to (HOMO-1)-LUMO. All values are reported in atomic units $a_0^{-2} \left(\hbar/E_h\right)^{-2} E_h^{-1}$.}
\label{tab:HL-amp-aprox}
\begin{tabular*}{\linewidth}{@{\extracolsep{\fill}}l *{4}{S[round-mode=places,round-precision=2,table-format=3.2]}}
\hline\hline
 & \multicolumn{2}{c}{\textbf{DHF}} & \multicolumn{2}{c}{\textbf{BH\&HLYP}} \\
\cline{2-3}\cline{4-5}
 & $\ensuremath{^1 \left(m_{hl}\right)_\mathrm{iso}}$ & {Eq. \eqref{eq:HL-amp-aprox}} & $\ensuremath{^1 \left(m_{hl}\right)_\mathrm{iso}}$ & {Eq. \eqref{eq:HL-amp-aprox}} \\
\hline
Zn$_2^{2+}$ & 1552,9084 & 1441,1126 & 3946,66286666667 & 2225,21146205 \\
Cd$_2^{2+}$ & 5964,81646666667 & 5362,1417 & 14361,2963333333 & 9251,35437587 \\
Hg$_2^{2+}$ & 77323,957 & 80495,4745 &  &  \\
\hline
Zn$_3^{2+}$ & 639,64974 & 704,221255864245 & 1506,55426666667 & 1019,75295950853 \\
Cd$_3^{2+}$ & 2505,7785 & 2754,32366187243 & 5545,9667 & 4097,00165028019 \\
Hg$_3^{2+}$ & 31603,3596666667 & 37451,7961529154 & 58374,679 & 54682,0037143287 \\
\hline\hline
\end{tabular*}
\end{table}

QED effects are taken into consideration by means of the following rescaling of the hfs integrals 
\begin{equation}
\label{eq:QED-corr}
\langle V \rangle_{ns} \to \langle V \rangle_{ns}(1 + \nu_{ns}^\mathrm{QED})
\end{equation}
where the factors ${\nu_{ns}^\mathrm{QED}}$ are taken from table II of Ref. \citenum{Sapirstein_PRA2003}. In order to obtain the corresponding $\nu_{ns}^\mathrm{QED}$ factors for zinc, cadmium and mercury, linear interpolation was applied to the values of Ref. \citenum{Sapirstein_PRA2003}. With the use of eqs. \eqref{eq:HL-amp-aprox} and \eqref{eq:QED-corr}, the corrected HOMO-LUMO amplitude takes the form
\begin{align}
\left(m_{hl}^\mathrm{QED}\right)_{\mu \mu} & \approx 2 \gamma \frac{\left(c_0 \langle V \rangle_{ns}\right)^2}{\epsilon_l - \epsilon_h} (1 + \nu_{ns}^\mathrm{QED})^2 \nonumber \\
\label{eq:HL-amp-aprox-QED-PZOA}
& = \left(m_{hl}\right)_{\mu \mu} + \delta^\mathrm{QED}
\end{align}
where the QED correction term is
\begin{equation}
\label{eq:QED-corr-PZOA}
\delta^\mathrm{QED} = 2 \gamma \frac{\left(c_0 \langle V \rangle_{ns}\right)^2}{\epsilon_l - \epsilon_h} \left[ 2\nu_{ns}^\mathrm{QED} + \left( \nu_{ns}^\mathrm{QED} \right)^2 \right]
\end{equation}

Since eqs. \eqref{eq:HL-amp-aprox} and \eqref{eq:HL-amp-aprox-QED-PZOA} apply for all values of $\mu = x,y,z$, the correction term $\delta^\mathrm{QED}$ is the same for the three diagonal components $\left(m_{hl}\right)_{xx}, \left(m_{hl}\right)_{yy}, \left(m_{hl}\right)_{zz}$ at PZOA. The total corrected diagonal components can be written, according to eq. \eqref{eq:J-PZOA}, as
\begin{equation}
\left(^1 {\bm J}^\mathrm{QED}\right)_{\mu \mu}^\mathrm{PZOA} = {^1 {\bm J}_{\mu \mu}^\mathrm{PZOA}} + \delta^\mathrm{QED}
\end{equation}
and the total isotropic component results
\begin{equation}
\label{eq:iso-QED-PZOA}
\left(^1 {\bm J}^\mathrm{QED}\right)_\mathrm{iso}^\mathrm{PZOA} = {^1 {\bm J}_\mathrm{iso}^\mathrm{PZOA}} + \delta^\mathrm{QED}
\end{equation}

At RPA level we must use the more general eq. \eqref{eq:bPb-reduced}
\begin{equation}
\label{eq:J-RPA}
{\bm J}_{\mu \nu}(M,N) = 2 \Re \left( \sum_{ia} {\left({\bm b}_{\mu}^K\right)_{ia}} ~ {\left({\bm X}_{\nu} ~ ^L\right)_{ia}^*} \right)
\end{equation}
For this, we multiply by unity so that each RPA diagonal component $\left(m_{hl}\right)_{\mu \mu}$ can be written in terms of its corresponding PZOA value
\begin{equation}
\label{eq:HL-amp-RPA}
\left(m_{hl}^\mathrm{RPA}\right)_{\mu \mu} = \frac{\left(m_{hl}^\mathrm{RPA}\right)_{\mu \mu}}{{\left(m_{hl}^\mathrm{PZOA}\right)_{\mu \mu}}} {\left(m_{hl}^\mathrm{PZOA}\right)_{\mu \mu}} = \mathbb{R}_{\mu} \cdot {\left(m_{hl}^\mathrm{PZOA}\right)_{\mu \mu}}
\end{equation}
Then, according to eqs. \eqref{eq:HL-amp-aprox-QED-PZOA} and \eqref{eq:HL-amp-RPA}, the corrected RPA HOMO-LUMO amplitude takes the form
\begin{align}
\left(m_{hl}^\mathrm{QED}\right)_{\mu \mu}^\mathrm{RPA} & = \mathbb{R}_{\mu} \cdot \left(m_{hl}^\mathrm{QED}\right)_{\mu \mu}^\mathrm{PZOA} \nonumber \\
\label{eq:HL-amp-aprox-QED-RPA}
& = \left(m_{hl}^\mathrm{RPA}\right)_{\mu \mu} + \mathbb{R}_{\mu} \delta^\mathrm{QED}
\end{align}
The correction to the total response function is, thus, included in a similar fashion at both PZOA and RPA, the only difference being that, at RPA, the QED correction $\delta^\mathrm{QED}$ is scaled by the corresponding factor $\mathbb{R}_{\mu}$. The total corrected diagonal components can, then, be written
\begin{equation}
\left(^1 {\bm J}^\mathrm{QED}\right)_{\mu \mu}^\mathrm{RPA} = {^1 {\bm J}_{\mu \mu}^\mathrm{RPA}} + \mathbb{R}_{\mu} \delta^\mathrm{QED}
\end{equation}
and the corresponding isotropic value results
\begin{align}
\label{eq:iso-QED-RPA}
\left(^1 {\bm J}^\mathrm{QED}\right)_\mathrm{iso}^\mathrm{RPA} = {^1 {\bm J}_\mathrm{iso}^\mathrm{RPA}} + \left( \frac{1}{3} \sum_{\mu} \mathbb{R}_{\mu} \right) \delta^\mathrm{QED}
\end{align}
%
\subsection{Computational details}\label{sec:comp-det}
%
Four-component calculations of nuclear spin-spin coupling were performed by means of the DIRAC code release 2019. \cite{DIRAC19} The Gaussian nuclear charge distribution was used \cite{Visscher1997} and the dyall.cv3z basis set was employed for describing the electronic structure of zinc,\cite{Dyall-unp} cadmium\cite{Dyall2007} and mercury\cite{Dyall2004} in all cases. Uncontracted Gaussian basis sets were used with the common gauge-origin (CGO) approach. The small component basis sets were generated by means of the unrestricted kinetic balance prescription (UKB). The complete set of calculations can be splitted into two main groups, one of which is based on the Dirac--Hartree--Fock (DHF) wave function, whilst the other was obtained by means of the density functional theory, with the hybrid BH\&HLYP (0.5 HF, 0.5 Slater, 0.5 Becke, 1 LYP) functional. In both cases, the Dirac--Coulomb Hamiltonian was used. Gaunt-type as well as (SS$\mid$SS) small-component two-electron integrals were neglected in all cases in order to reduce the computational cost. Test calculations on the model system Hg$_3^{2+}$ with dyall.cv3z basis set revealed that the one-bond coupling $^1 J_{zz}$ was lowered by as much as 0.3\% by the inclusion of (SS$\mid$SS) two-electron integrals and further inclusion of Gaunt integrals represents a lowering by an amount of about 0.65\% (both for DHF and BH\&HLYP wave functions). The use of dyall.cv4z basis set (without (SS$\mid$SS) and Gaunt integrals) lowers the dyall.cv3z value by $\sim$ 0.4\% for DHF and by $\sim$ 0.3\% for BH\&HLYP. Similar comparisons for the two-bond coupling $^2 J_{zz}$ value in these preliminary calculations reveal a departure of, at most, 1.4\% in all cases. Non-relativistic values of spin-spin couplings were obtained by rescaling the speed of light as $c = 100c_0$. Linear response calculations were performed within the relativistic polarization propagator approach at the RPA and PZOA levels of theory.

The geometries of all systems X$_2^{2+}$ and X$_3^{2+}$ (X = Zn, Cd, Hg) were obtained after optimization with Dirac--Coulomb Hamiltonian and DHF wave function and using dyall.cv3z basis set for the atomic centers. The final internuclear distances are presented in the Supplementary Information.

Expansion coefficients in eqs. \eqref{eq:HOMO} and \eqref{eq:LUMO} were obtained with the use of \textit{projection analysis}\cite{Dubillard_JCP2006, Saue_JCP2020} combined with \textit{intrinsic atomic orbital (IAO)} approach.\cite{Knizia_JCTC2013} For this, we first obtained the four-component SCF wave function for each atomic center X, saving the corresponding coefficients. Then, the molecular wave function was obtained for X$_2^{2+}$ (or X$_3^{2+}$), saving its coefficients too. Finally, the projection analysis was performed by means of saved coefficients, both atomic and molecular. This function allows to define the set of atomic orbitals $\{ \chi_p^A, \chi_{\bar{p}}^A \}$ onto which the projection is to be made. The choice of reference orbitals consisted on all occupied atomic orbitals of the fragments in their ground states. These correspond to the following orbital strings: $1,\dots,15$ for zinc, $1,\dots,24$ for cadmium and $1,\dots,40$ for mercury.

However, the above mentioned choice of reference orbitals did not provide a satisfactory description of virtual molecular orbitals beyond LUMO, due to the large contribution of the polarization term (see eqs. \eqref{eq:HOMO} and \eqref{eq:LUMO}). While the inclusion of virtual atomic orbitals associated to each fragment in the set of reference orbitals naturally reduced the polarization contribution, this prescription also led to projection coefficients larger than unity in the decomposition of virtual molecular orbitals, making the analysis out of scale. These subtleties made us settle with the aforementioned sets of reference orbitals and limit our approach just to LUMO (which was the virtual molecular orbital with lowest polarization contribution in all cases).

Quasi-relativistic two-component calculations within zeroth order
regular approximation (ZORA) were performed with a modified
version\cite{wullen:2010} of the program package
turbomole.\cite{ahlrichs:1989} ZORA calculations were
performed within complex generalized Hartree-Fock (cGHF) or Kohn-Sham
(cGKS). In all these calculations we employed the dyall.cv3z basis set.
Picture-change transformed $J$-coupling tensors were computed with our
toolbox approach detailed in Ref.~\citenum{gaul:2020}. For
optimization of the response function we used the approach derived in
Ref.~\citenum{bruck:2021}. Uncoupled (UC) results were obtained
following eq. (31) of Ref.~\citenum{gaul:2020}, whereas coupled
perturbed (CP) results were obtained after self-consistent optimization of 
the uncoupled orbital rotation matrix as described in Sec. II a of
Ref.~\citenum{bruck:2021}. We expanded the implementation of
Ref.~\citenum{bruck:2021} to open-shell systems and GGA functionals.
The extension to open-shell systems is straightforward as all
equations of Ref.~\citenum{bruck:2021} apply, and in addition, just
the operator is allowed to be complex spin-independent or real
spin-dependent (for details see Ref.~\citenum{gaul:2020}). For the
exchange-correlation potential the extension to open-shell systems and
GGA functionals follows as shown in Appendix B of
Ref.~\citenum{bruck:2021} and Ref.~\citenum{wullen:2007}. The explicit
functional derivatives
$\frac{\delta
\hat{V}_\mathrm{XC}^{\kappa}[\{\rho_\mathrm{e}^\lambda(r)\},\{\vec{\nabla}\rho_\mathrm{e}^\lambda(r)\}]}{\delta\rho_\mathrm{e}^\lambda(r)}$
and
$\frac{\delta\hat{V}_\mathrm{XC}^{\kappa}[\{\rho_\mathrm{e}^\lambda(r)\},\{\vec{\nabla}\rho_\mathrm{e}^\lambda(r)\}]}{\delta\vec{\nabla}\rho_\mathrm{e}^\lambda(r)}$
are given elsewhere.

%
\section{Results and discussions}\label{sec:results}
%

When presenting the results, we shall use, rather than the $\bar{\bar{J}}$ tensor, the {\it reduced indirect nuclear spin-spin coupling tensor} $\bar{\bar{K}}$ whose components are given (in SI units $\mathrm{N} \cdot \mathrm{m}^{-3} \cdot \mathrm{A}^{-2}$) by
\begin{equation}
K_{\mu\nu}(M,N) = \frac{4\pi^2}{h\gamma_M \gamma_N} J_{\mu\nu}(M,N).
\end{equation}
Since it is independent of magnetic moments $\gamma_M$, $\gamma_N$ of
the coupled nuclei, it presents the advantage of being (apart from
finite nuclear size effects) nearly
independent of different isotopes that may be involved in a given ion
and thus is determined mainly by
the value of the response function.

In the calculation of one-bond couplings in X$_3^{2+}$ ions, the main coupling pathway is that involving ``HOMO-1'' (the molecular orbital energetically just below HOMO) and LUMO so, strictly speaking, it is the $\left(m_{h-1,l}\right)_{\mu \mu}$ amplitude that needs to be separated from the sum in eq. \eqref{eq:J-PZOA}, $\ket{\phi_{h-1}}$ the molecular orbital that is to be considered in eq. \eqref{eq:HOMO} and the energy gap $\epsilon_l - \epsilon_{h-1}$ that is to be used in eq. \eqref{eq:HL-amp-aprox}; the rest of the treatment developed in section \ref{sec:model} is equally applied after these considerations. That the $\left(m_{h-1,l}\right)_{\mu \mu}$ amplitude is the most important in one-bond couplings in X$_3^{2+}$ ions is closely related to the fact that projection analysis reveals no participation in the HOMO of atomic orbitals associated to the central nucleus as is expected from this essntially non-bonding molecular orbital.

The rest of this section is organized as follows. In section \ref{sec:J-Hg}, we start with the case of spin-spin couplings in the model mercury ions Hg$_2^{2+}$ and Hg$_3^{2+}$, where a comparison with other published (calculated and measured) results is made. In section \ref{sec:total} total values for the one-bond coupling and its main contributing amplitudes are analyzed for X$_3^{2+}$ and X$_2^{2+}$ ions. A similar approach is followed within the next two sections: in section \ref{sec:rel-eff} the relativistic effects are analyzed and compared with non-relativistic contributions for DHF and BH\&HLYP wave functions and in section \ref{sec:QED-eff} a similar analysis is developed for QED corrections. Finally, in section \ref{sec:QED-rel-tot} we end up with the analysis of QED corrections when compared with the relativistic effects and total one-bond couplings.

%
\subsection{Spin-spin coupling in mercury ions Hg$_2^{2+}$ and Hg$_3^{2+}$}\label{sec:J-Hg}
%

In Table \ref{tab:Hg} we collect calculated values for the isotropic
component of the reduced spin-spin coupling tensor $\bar{\bar{K}}$ (in $10^{21} ~ \mathrm{N} \cdot \mathrm{m}^{-3} \cdot \mathrm{A}^{-2}$) for the mercury ions
Hg$_2^{2+}$ and Hg$_3^{2+}$, together with the results of Refs.
\citenum{Autschbach_JACS2003, Autschbach_JACS2007} and available
experimental data. As a first remark, we note that at the
level of PZOA there
exists a lowering of the coupling with an increasing number of bonds, both for $c = c_0$ and $c = 100c_0$ (this can be observed in the
fact that ${^{1}K}(\text{Hg}_3^{2+}) > {^{2}K}(\text{Hg}_3^{2+})$), as
is expected, at least, in the non-relativistic regime. However, this
trend breaks when we consider the RPA, both for $c = c_0$
and $c = 100c_0$.
This fact is notably accentuated for the relativistic Dirac-Coulomb
value, where ${^{2}K}(\text{Hg}_3^{2+})$ surpasses
${^{1}K}(\text{Hg}_3^{2+})$ by a factor $\sim$ 4.6. This contrasts
with the corresponding BH\&HLYP value, for which
${^{2}K}(\text{Hg}_3^{2+})$ is $\sim$ 2.4 times greater than
${^{1}K}(\text{Hg}_3^{2+}) ~ (c = c_0)$. A similar calculation using the
hybrid functional PBE0 reproduces this result with a rough factor of
2. This indicates that the electronic correlation, as described by the
density functional theory, has a sizeable effect on the coupling. 
To investigate this further, we studied different functionals
on the level of ZORA and find that the ratio of
${^{2}K}(\text{Hg}_3^{2+})/{^{1}K}(\text{Hg}_3^{2+})$ decreases
when reducing the amount of exact Fock exchange and when going
from GGA to LDA.

As a
further remark, the non-relativistic RPA values on Hg$_3^{2+}$ reveal
that the ratio ${^{2}K}(\text{Hg}_3^{2+})/{^{1}K}(\text{Hg}_3^{2+})
\approx 3/2$, a result that is also predicted by the Hückel model,
when applied for the description of the valence molecular orbitals in
the Hg$_3^{2+}$ system, taking the atomic $6s$ orbitals centered in
each mercury nucleus as basis functions and using the Pople--Santry
model  for the description of the spin-spin
coupling.\cite{Autschbach_JACS2003}

As is known, calculations performed on bare ions in gas phase, as in
this work, do not necessarily reproduce experimental results
obtained in condensed phases, but this can be cured by
taking into account solvent effects and polarization of the electron
density.\cite{Autschbach_JACS2003, Autschbach_JACS2007} In fact, the
increment of the coupling with the internuclear distance was also
observed by Autschbach and co-workers in the bare Hg$_3^{2+}$ ion,
using the ZORA Hamiltonian.  Comparison to our
present ZORA results with the same functionals (LDA and BP86) gives
reasonable agreement with maximum deviations of 6~\% on ${^{2}K}$ and
10~\% on ${^{1}K}$ in $\text{Hg}_3^{2+}$. The deviations can be
explained primarilly by the use of molecular structures with very
different bond lengths and, additionally, by the use of a different type
of basis set (Gaussian type basis set in the present work instead of
Slater type functions as used in Refs.~\citenum{Autschbach_JACS2003,
Autschbach_JACS2007}). The deviations on ${^{1}K}$ in
$\text{Hg}_2^{2+}$ are $6~\%$ on the level of BP86 but are slightly
larger on the LDA level. Furthermore, in all our calculations, as well
as in the results presented in Ref.~\citenum{Autschbach_JACS2003} on $\text{Hg}_3^{2+}$, we
see a trend that the magnitude of absolute values of
${^{1}K}$/${^{2}K}$ decreases from HF via hybrid DFT and GGA to LDA,
whereas the LDA value for $\text{Hg}_2^{2+}$ in
Ref.~\citenum{Autschbach_JACS2003} contradicts this trend as it is
larger than the value that was obtained with BP86 in this work.

\begin{table*}[ht!]
\setlength{\extrarowheight}{1.1mm}
\caption{Isotropic components of reduced coupling tensor $K_\mathrm{iso}$ (in $10^{21} ~ N \cdot m^{-3} \cdot A^{-2}$) in $\text{Hg}_2^{2+}$ and $\text{Hg}_3^{2+}$. }
\label{tab:Hg}
\begin{tabular*}{\linewidth}{@{\extracolsep{\fill}}ll *{3}{S[round-mode=figures,round-precision=3]}}
\hline\hline
& & {$^{1}K(\text{Hg}_2^{2+})$} & {${^{1}K}(\text{Hg}_3^{2+})$} & {${^{2}K}(\text{Hg}_3^{2+})$} \\
\hline
\multicolumn{2}{l}{This work:} &&&\\
\multirow{9}{*}{PZOA/UC} &4c-DHF $(c = 100 c_0)$ & 350.569151747797 & 162.655552 & 76.288961 \\
&4c-DHF $(c = c_0)$ & 2840.73265664194 & 1253.855617 & 831.741350 \\
&ZORA-HF         &    2831.28 & 1247.36 &  826.05\\
&4c-DKS-BH\&HLYP $(c = 100 c_0)$ & 749.137742 & 307.115563 & 218.422568 \\
&4c-DKS-BH\&HLYP $(c = c_0)$ & 5716.315239 & 2217.21739 & 2187.782414 \\
&ZORA-BH\&HLYP   &    5672.75 & 2199.83 & 2167.35 \\
&ZORA-PBE0       &    8615.75 & 3059.47 & 3702.62 \\
&ZORA-BP86       &   15384.20 & 4689.20 & 7522.27 \\
&ZORA-LDA        &   15506.70 & 4711.52 & 7603.00 \\
\\
\cline{2-5}
\\
\multirow{10}{*}{RPA/CP} & 4c-DHF $(c = 100c_0)$ & 3178.81526977504 & 965.936843 & 1438.178604 \\
&4c-DHF $(c = c_0)$ & 67851.9155544773 & 8713.412116 & 39982.593570 \\
&ZORA-HF         &   66786.90 & 8633.57 & 39250.9\\
&4c-DKS-BH\&HLYP $(c = 100c_0)$ & 2809.464695 & 835.926697 & 1124.653301 \\
&4c-DKS-BH\&HLYP $(c = c_0)$ & 32043.060515 & 6614.675381 & 15827.101737 \\
&ZORA-BH\&HLYP   &   32037.90 & 6492.89 & 15302.80\\
&4c-DKS-PBE0 $(c = c_0)$ & 26389.929880 & 6141.321378 & 13043.629259 \\
&ZORA-PBE0       &   25376.00 & 6045.18 & 12698.30\\
&ZORA-BP86       &   21946.80 & 5584.65 & 10984.40\\
&ZORA-LDA        &   19488.20 & 5289.70 &  9545.59 \\
\hline
\multicolumn{2}{l}{Other theory: }&&&\\
\\
\multicolumn{2}{l}{Autschbach et al.\cite{Autschbach_JACS2003}$^a$, sc-ZORA (PBE) }&  & 6599.550928 & 12549.54975 \\
\multicolumn{2}{l}{Autschbach et al.\cite{Autschbach_JACS2003}$^a$, so-ZORA (PBE) }&  & 6584.327051 & 12349.10203 \\
\multicolumn{2}{l}{Autschbach et al.\cite{Autschbach_JACS2003}$^a$, sc-ZORA (PBE - COSMO) }&  & 5838.35705 & 10108.65471 \\
\multicolumn{2}{l}{Autschbach and Sterzel\cite{Autschbach_JACS2007}$^b$, sc-ZORA (GGA - BP) }& 23744.17439 & 6135.222662 & 11681.78873 \\
\multicolumn{2}{l}{Autschbach and Sterzel\cite{Autschbach_JACS2007}$^b$, so-ZORA (GGA - BP) }& 20800.89139 & 6114.924159 & 11483.87832 \\
\multicolumn{2}{l}{Autschbach and Sterzel\cite{Autschbach_JACS2007}$^b$, sc-ZORA (LDA - VWN) }& 23888.80123 & 6163.133105 & 11775.6693 \\
\multicolumn{2}{l}{Autschbach and Sterzel\cite{Autschbach_JACS2007}$^b$, so-ZORA (LDA - VWN) }& 20927.75704 & 5823.133172 & 10093.43083 \\
\hline
\multicolumn{2}{l}{Experiment: }&&&\\
\\
\multicolumn{2}{l}{Malleier et al.\cite{Malleier_CC2001} }& 7208.506031 &  &  \\
\multicolumn{2}{l}{Gillespie et al.\cite{Gillespie_IC1984} }&  & 3544.626162 &  \\
\hline\hline
\end{tabular*}

\smallskip

$^a$ Table 1 of Ref. \citenum{Autschbach_JACS2003} (coupling between $^{199}$Hg nuclei). $R_0(\text{Hg-Hg}) = 2.638$~\AA\ in Hg$_2^{2+}$ and $R_0(\text{Hg-Hg}) = 2.665$~\AA\ in Hg$_3^{2+}$. $^b$ Table 1 of Ref. \citenum{Autschbach_JACS2007}. $R_0(\text{Hg-Hg}) = 2.776$~\AA\ for the first two cases (optimized in gas phase);  $R_0(\text{Hg-Hg}) = 2.670$~\AA\ for the third case (optimized using the COSMO model). 
\end{table*}

%
\subsection{Total spin-spin couplings and principal coupling pathway contributions}\label{sec:total}
%

In the following discussion, we refer exclusively to the main coupling amplitudes that contribute to the ``paramagnetic'' or ($e$-$e$) part of the response function, namely HOMO-LUMO for X$_2^{2+}$ and (HOMO-1)-LUMO for X$_3^{2+}$. We found that ($e$-$e$) contribution surpasses 99.99\% of the total isotropic component of the coupling tensor at PZOA and 99.98\% at RPA, in all cases (i.e., DHF and BH\&HLYP calculations, both with $c = c_0$ and $c = 100c_0$).

%
\subsubsection{One-bond couplings in X$_3^{2+}$}
%
From table \ref{tab:total-values}, we see that there are different patterns of (HOMO-1)-LUMO contributions at PZOA and RPA levels. At PZOA, DHF values increment monotonically from 84.2\% (Zn$_3^{2+}$) to 90.6\% (Hg$_3^{2+}$) for $c = c_0$, but diminish from 81.9\% (Zn$_3^{2+}$) to 73\% (Hg$_3^{2+}$) for $c = 100c_0$. A similar behaviour can be observed with BH\&HLYP results for which (HOMO-1)-LUMO contribution exhibits a slight increment from 92.7\% (Zn$_3^{2+}$) to 94.6\% (Hg$_3^{2+}$) for $c = c_0$, but diminishes from 91\% (Zn$_3^{2+}$) to 83.6\% (Hg$_3^{2+}$) for $c = 100c_0$.

Analogously, at RPA, DHF results show a monotonic increment of (HOMO-1)-LUMO contribution from 50.9\% (Zn$_3^{2+}$) to 60.1\% (Hg$_3^{2+}$) for $c = c_0$, but remains about 50\% in all three cases for $c = 100c_0$. These are the lowest amounts in which (HOMO-1)-LUMO coupling pathway contributes to the total value of all cases considered in this work. However, the corresponding BH\&HLYP results show a monotonic increment from 70.8\% (Zn$_3^{2+}$) to 77.2\% (Hg$_3^{2+}$) for $c = c_0$, but this is reversed for $c = 100c_0$, for which 69.5\% is observed in Zn$_3^{2+}$, until it reaches 67.9\% in Hg$_3^{2+}$.

\begin{figure}[h!]
\centering
\includegraphics[scale=0.7]{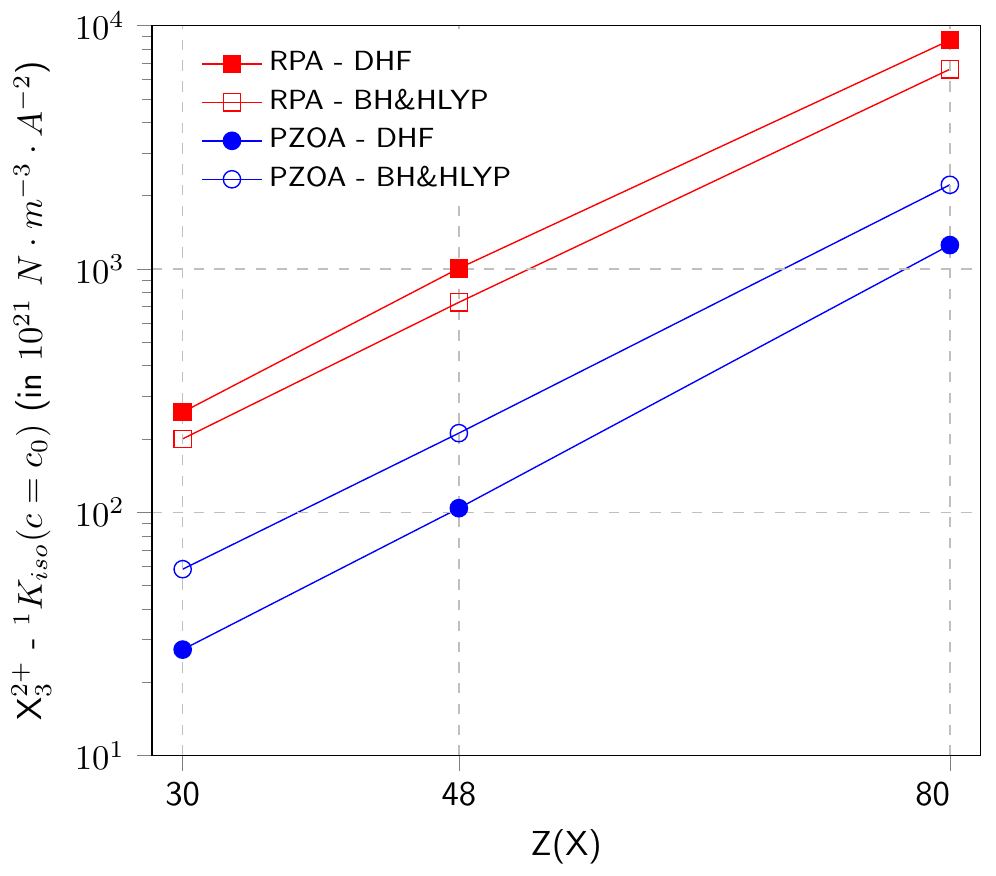}
\caption{Total values of one-bond isotropic reduced coupling $^1 K_\mathrm{iso}$ (in $10^{21} ~ \mathrm{N} \cdot \mathrm{m}^{-3} \cdot \mathrm{A}^{-2}$) in X$_3^{2+}$ ions, as a function of the atomic number of element X, Z(X). PZOA values are plotted with blue circular marks and RPA ones, with red squared marks. DHF results are identified by  filled marks and the corresponding BH\&HLYP, with hollow ones. A logarithmic scale was set for the vertical axis, thus indicating a power-law according to $Z^4$.}
\label{fig:K1-X3-PZOA-RPA}
\end{figure}

Total values of the isotropic component of the reduced one-bond coupling $^1 K_\mathrm{iso}$ in X$_3^{2+}$ ions are plotted in Fig.\ref{fig:K1-X3-PZOA-RPA}, as a function of the atomic number of element X, $Z$(X), where we compare PZOA and RPA values, obtained with DHF and BH\&HLYP wave functions (for $c = c_0$). The gap between PZOA-DHF and RPA-DHF values reaches almost one order of magnitude in a given ion, a difference that is notably reduced when the BH\&HLYP functional is used. We also note a characteristic power-law as a function of $Z$(X) in all sets of values. To test this observed trend a linear regression analysis was performed based on the functional form $\ln Q = \ln m + n\ln Z$ (which is the linearized equivalent of the power-law $Q = mZ^n$). We show the corresponding parameters in Table \ref{tab:regression}, where $Q$ stands for various quantities of interest (among them, total values of one-bond coupling $^1 K_\mathrm{iso}(c = c_0)$ and $^1 K_\mathrm{iso}(c = 100c_0)$, together with QED corrections $|\delta^\mathrm{QED}(^1 K_\mathrm{iso})|$ and
relativistic effects $\delta^\mathrm{Rel}(^1 K_\mathrm{iso})$). The regression analysis suggests a power-law of the form $^1 K_\mathrm{iso}(\text{X}_3^{2+}) \propto Z^4$. In X$_3^{2+}$ ions, this trend is clearer when PZOA values are used. This can be seen as a consequence of the predominant (HOMO-1)-LUMO path, $\left(m_{h-1,l}\right)$ at PZOA level (as mentioned above), which is dominated by the contribution of the valence $s$-type atomic orbitals of the constituents. Taking into account eq. \eqref{eq:HL-amp-aprox}, we note that this contribution is proportional to the square of the hfs integral $\left(\langle V \rangle_{ns}\right)^2$, for which we obtained a dependence of the form $Z^2$ (see Table \ref{tab:regression}), thus yielding the mentioned functional form for $\left(m_{h-1,l}\right)$ (the factor $\gamma$ and the energy gap $\epsilon_l - \epsilon_{h-1}$ remain nearly constant through the whole $Z$(X) series in all cases). In contrast, the higher contribution of terms other than $\left(m_{h-1,l}\right)$ at RPA level results in a lower expected value for the parameter $n \approx 3.6$.
%
\subsubsection{One-bond couplings in X$_2^{2+}$}
%
The same analysis can be made for the case of X$_2^{2+}$ ions, for wihch we find similar patterns to those described above for the HOMO-LUMO contribution in the various cases studied (see table \ref{tab:total-values}). For example, at PZOA the HOMO-LUMO term remains near the 95\% of the total response value in DHF results (both for $c = c_0$ and $c = 100c_0$). This percentage is even more pronounced with BH\&HLYP, for which HOMO-LUMO contribution is almost equal to the total response value (both for $c = c_0$ and $c = 100c_0$). At RPA, the HOMO-LUMO amplitude represents a lower proportion of the total value, but still remains above the 71\% for DHF ($c = c_0$ and $c = 100c_0$) and is about 85\% for BH\&HLYP ($c = c_0$ and $c = 100c_0$).

A plot, similar to Fig. \ref{fig:K1-X3-PZOA-RPA}, for the case of X$_2^{2+}$ ions is given in the Supplementary Information. It shows a similar behaviour to that observed for X$_3^{2+}$ ions. Total values of one-bond coupling are notoriously raised in X$_2^{2+}$ ions, when compared with the X$_3^{2+}$ case, as is the difference between PZOA-DHF and RPA-DHF as well. The use of the density functional has the same effect as that described in the preceding section.

Linear regression was performed for this case too and reveals a power-law of the form $^1 K_\mathrm{iso}(\text{X}_2^{2+}) \propto Z^4$ (see Table \ref{tab:regression}), which can be explained by a similar argument as in the previous subsection.
\begin{table*}[!t]
\caption{Isotropic ($ee$) values for total one-bond reduced coupling tensor, $^1 K_\mathrm{iso}$, and its main contributing amplitudes: (HOMO-1)-LUMO (X$_3^{2+}$) and HOMO-LUMO (X$_2^{2+}$). The left half of the table collects DHF values, divided into relativistic ($c = c_0$) and non-relativistic ($c = 100c_0$) results. The same applies for the right half and BH\&HLYP. For a given ion, the table splits into PZOA and RPA approximations. All values are presented in units of $10^{21} N \cdot m^{-3} \cdot A^{-2}$.}
\label{tab:total-values}
\setlength{\extrarowheight}{1.1mm}
\begin{tabular*}{\linewidth}{@{\extracolsep{\fill}}c l  *{4}{S[round-mode=places,round-precision=3]}  *{4}{S[round-mode=places,round-precision=3]}}
\hline
\hline
 &  & \multicolumn{4}{c}{\textbf{DHF}} & \multicolumn{4}{c}{\textbf{BH\&HLYP}} \\
\cline{3-6}\cline{7-10}
 & & \multicolumn{2}{c}{$c = c_0$} & \multicolumn{2}{c}{$c = 100c_0$} & \multicolumn{2}{c}{$c = c_0$} & \multicolumn{2}{c}{$c = 100c_0$} \\
\cline{3-4}\cline{5-6}\cline{7-8}\cline{9-10}
 & X & Total & $m_{hl}$ & Total & \multicolumn{1}{c}{$m_{hl}$} & \multicolumn{1}{c}{Total} & $m_{hl}$ & Total & $m_{hl}$ \\
\hline
\multicolumn{1}{c}{} & \multicolumn{1}{c}{} & \multicolumn{8}{c}{$^{1}\mathbf{K}(\textbf {X}_2^{2+})$} \\
\hline
\multirow{3}{1.3cm}{\textbf{PZOA}} & Zn & 58,4568832359467 & 55,8049985362105 & 45,9766936338113 & 43,3511072505366 & 141,253370983486 & 141,826469286437 & 110,991142189025 & 110,823614061699 \\
 & Cd & 220,279771164947 & 214,35042414034 & 114,480736169026 & 107,937703766251 & 510,584889880993 & 516,084606702979 & 262,437910111818 & 261,379907429348 \\
 & Hg & 2840,73265664194 & 2778,69789821409 & 350,569151747797 & 309,890079649108 & 5716,31522773873 & 5694,76274610439 & 749,137742436235 & 715,974617704993 \\
\hline
\multirow{3}{1.3cm}{\textbf{RPA}} & Zn & 1314,33544505663 & 942,367948615484 & 856,281035540666 & 609,667023906764 & 789,457302538609 & 674,149436562317 & 578,209562778948 & 491,673959925184 \\
 & Cd & 6150,45644319553 & 4483,10487167451 & 1905,0667443824 & 1366,87858971954 & 3070,01936713506 & 2648,98244600868 & 1309,02923481327 & 1117,92294571042 \\
 & Hg & 67851,9155544773 & 51758,3560600673 & 3178,81526977504 & 2275,51131928142 & 32043,0604515531 & 28068,0419416934 & 2809,46468612232 & 2381,90824946701 \\
\hline
\multicolumn{1}{c}{} & \multicolumn{1}{c}{} & \multicolumn{8}{c}{$^{1}\mathbf{K}(\textbf{X}_3^{2+})$} \\
\hline
\multirow{3}{1.3cm}{\textbf{PZOA}} & Zn & 27,3036300516239 & 22,9863221838374 & 21,5562667680384 & 17,6501535165015 & 58,4287660899327 & 54,1392258848333 & 46,1974252772699 & 42,0399106966443 \\
 & Cd & 104,051464839524 & 90,0471434918941 & 54,7786815094057 & 44,2718314409196 & 211,632726454287 & 199,298724622375 & 111,567995149621 & 100,247522609557 \\
 & Hg & 1253,85561463861 & 1135,69186691093 & 162,655553982086 & 118,681506377662 & 2217,2173859043 & 2097,74052104217 & 307,115558754949 & 256,887361625292 \\
\hline
\multirow{3}{1.3cm}{\textbf{RPA}} & Zn & 258,169207604708 & 131,481264121875 & 197,026157648885 & 98,2535207880217 & 200,469949929456 & 141,820033185481 & 155,691644967726 & 108,151155203764 \\
 & Cd & 1006,356510829 & 528,632452125724 & 477,235805969903 & 235,877726438479 & 729,419104322905 & 529,795310507647 & 367,567752734368 & 254,289736665557 \\
 & Hg & 8713,4125887346 & 5240,24153531107 & 965,930914217971 & 485,840777765785 & 6614,67423303288 & 5103,65388929389 & 835,942197464281 & 567,963892703864 \\
\hline
\hline
\end{tabular*}
\end{table*}
%
\subsection{Relativistic effects on total isotropic coupling}\label{sec:rel-eff}
%
In the following, we take relativistic effects as given by the difference between a result obtained after a four-component calculation without modification of the speed of light ($c = c_0$) and the corresponding four-component result obtained by setting $c = 100c_0$.
This was done either for DHF or BH\&HLYP wave functions, and for both PZOA and RPA approximations. In this sense, for instance, the total relativistic effect on the isotropic one-bond reduced coupling is calculated as $\delta^\mathrm{Rel}(^1 K_\mathrm{iso}) = {^1 K_\mathrm{iso}} (c = c_0) - {^1K_\mathrm{iso}} (c = 100c_0)$ for any of the four cases just mentioned. The relativistic effect can, analogously, be defined for any related
quantity. When a given quantity has no explicit dependence on the value of $c$, we are referring to its relativistic value ({\it i. e.}, that obtained with $c = c_0$).
%
\subsubsection{One-bond couplings in X$_3^{2+}$}
\begin{figure}[h!]
\centering
\includegraphics[scale=0.7]{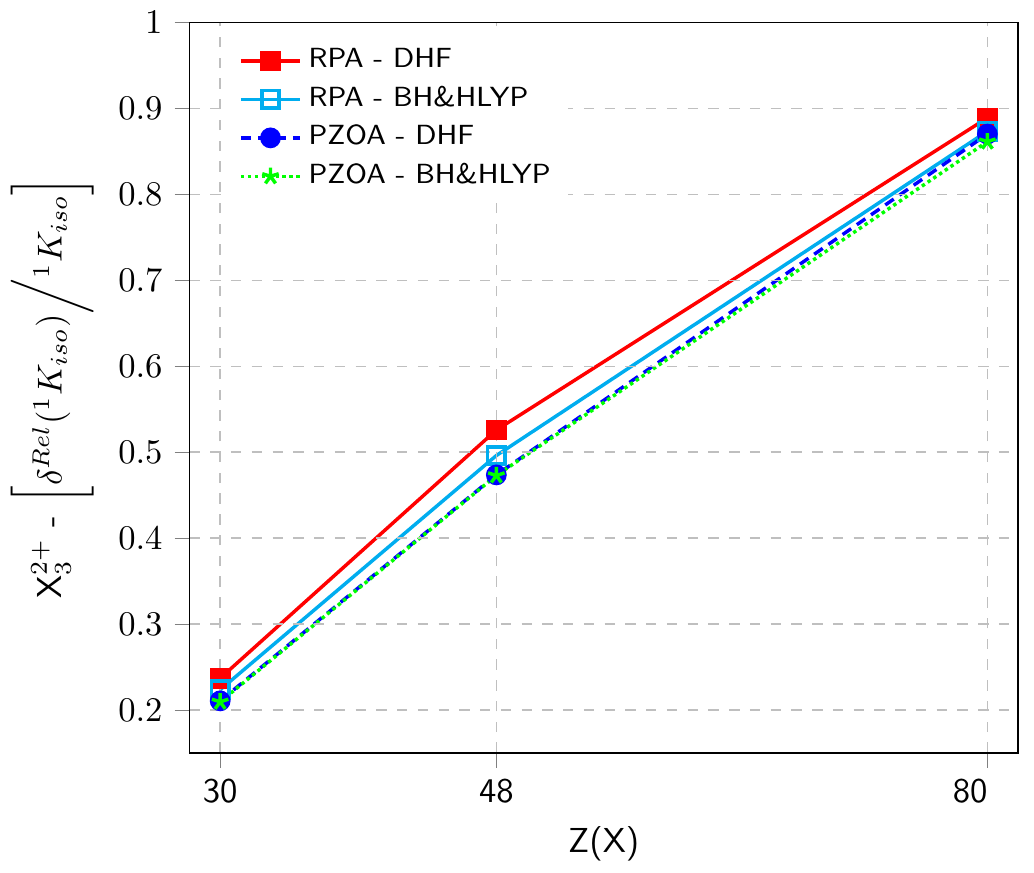}
\caption{Ratio of the relativistic effect to the total value of isotropic one-bond reduced coupling $\delta^\mathrm{Rel}(^1 K_\mathrm{iso})/{^1 K_\mathrm{iso}}$ as a function of the atomic number Z(X) in X$_3^{2+}$ ions. PZOA-DHF values are plotted with blue dashed line and filled circles and PZOA-BH\&HLYP with green dotted line and stars (the curves overlap in part). In red solid line with filled squares, the values for RPA-DHF and the cyan solid line with hollow squares, the RPA-BH\&HLYP ones.}
\label{fig:X3-I-rel-eff}
\end{figure}

%
In fig. \ref{fig:X3-I-rel-eff} the ratio of the relativistic effect to the total value of isotropic one-bond reduced coupling $\delta^\mathrm{Rel}(^1 K_\mathrm{iso})/{^1 K_\mathrm{iso}}$ is plotted as a function of the atomic number Z(X) for the various cases considered using X$_3^{2+}$ ions. We note that, within PZOA approximation, both cases present almost the same behaviour, specially for the lighter ions, and evidence a linear increment of the relative contribution of relativistic effects with $Z$(X). At RPA level, the results slightly depart from the linear trend. This departure reaches its maximum at Cd$_3^{2+}$, being larger for RPA-DHF values than for RPA-BH\&HLYP, indicating a greater contribution of relativity for DHF than BH\&HLYP.

In Hg$_3^{2+}$, the relativistic effects represent a significant amount of the total value, reaching almost a
90\% in all cases. Furthermore, we note that ZORA agrees with Dirac-Coulomb calculations within 5~\%\ indicating that valence electrons have a dominant role in the indirect nuclear spin-spin couplings, for which relativistic
effects are excellently described within the ZORA approach even for properties that are vanishing in the non-relativistic limit.\cite{berger:2005,nahrwold:09,gaul:2017,gaul:2020}

Regression analysis was also performed for the relativistic effect, according to the linearized form $\ln Q = \ln m + n\ln Z$ as in the previous section. The results (listed in Table \ref{tab:regression}) show a dependence $\delta^\mathrm{Rel}(^1 K_\mathrm{iso}) \propto Z^5$. The scaling of relativistic effects on one-bond couplings as $\sim Z^{4}$ is expected for heavy atoms from leading order relativistic contributions to the spin-spin couplings. For an overview of all relativistic corrections to leading order see chapter 13 of Ref.~\onlinecite{Kaupp2004} and the more recent review of Ref.~\onlinecite{GAA_IJQC2017}. Further studies needs to be performed to find the origin of the additional factor of $Z$ that is observed for the relativistic effects on total spin-spin coupling constants. 
%
\subsubsection{One-bond couplings in X$_2^{2+}$}
%
The corresponding results for X$_2^{2+}$ ions are presented in the Supplementary Information. We note the remarkable match between values obtained within PZOA; this fact does not only hold when using different wave functions for a given ion, but also extends when the comparison is made between X$_2^{2+}$ and X$_3^{2+}$ ions. The relativistic effects, thus, have essentially the same relative contribution to the one-bond coupling at PZOA level, whether described by DHF or BH\&HLYP wave functions and irrespective of the kind of ion that is considered. In X$_2^{2+}$ ions, the difference between RPA and PZOA become notoriously enlarged with respect to the X$_3^{2+}$ case and a more pronounced contribution of relativity is observed with RPA-DHF than with RPA-BH\&HLYP. In this sense, the relativistic effects almost equal the total value at Hg$_2^{2+}$, as described by RPA-DHF and the ZORA approach again recovers the RPA-DHF values within 5~\%.

We further note that a similar power-law $\delta^\mathrm{Rel}(^1 K_\mathrm{iso}) \propto Z^5$ is recovered for the X$_2^{2+}$ ions (see Table \ref{tab:regression}).
%
\subsection{QED corrections on total isotropic coupling}\label{sec:QED-eff}
%
\begin{table}[ht!]
\setlength{\extrarowheight}{1.5mm}
\caption{Calculated QED corrections $\delta^\mathrm{QED}$ (in $10^{21} ~ \mathrm{N} \cdot \mathrm{m}^{-3} \cdot \mathrm{A}^{-2}$) to the isotropic reduced one-bond coupling $^1 K_\mathrm{iso}$ within the various descriptions. For PZOA approximation, $\delta^\mathrm{QED}$ is calculated according to eq. \eqref{eq:iso-QED-PZOA}, whereas eq. \eqref{eq:iso-QED-RPA} is used for RPA approximation.}
\label{tab:QED-eff}
\begin{tabular*}{\linewidth}{@{\extracolsep{\fill}}l *{4}{S[round-mode=places,round-precision=2,table-format=3.2]}}
\hline\hline
 & \multicolumn{2}{c}{\textbf{DHF}} & \multicolumn{2}{c}{\textbf{BH\&HLYP}} \\
\cline{2-3}\cline{4-5}
 & {\textbf{PZOA}} & {\textbf{RPA}} & {\textbf{PZOA}} & {\textbf{RPA}} \\
\hline
Zn$_2^{2+}$ & -0,200868033360357 & -3,39223590364581 & -0,310158875236283 & -1,47432599538848 \\
Cd$_2^{2+}$ & -1,38924519083828 & -29,0583951215357 & -2,39687802772887 & -12,3031718574242 \\
Hg$_2^{2+}$ & -32,5382603174247 & -606,181535699001 &  &  \\
\hline
Zn$_3^{2+}$ & -0,098157175783901 & -0,561575461839761 & -0,142137246879592 & -0,37236319266298 \\
Cd$_3^{2+}$ & -0,71360123049832 & -4,19033430804433 & -1,06146763340299 & -2,82194003220884 \\
Hg$_3^{2+}$ & -15,1389416700476 & -69,872086416757 & -22,1038174311462 & -53,7817477543226 \\
\hline\hline
\end{tabular*}
\end{table}
%
\subsubsection{One-bond couplings in X$_3^{2+}$}
%
In Table \ref{tab:QED-eff} the values of QED corrections on the isotropic reduced one-bond coupling $delta^\mathrm{QED}(^1 K_\mathrm{iso})$ are presented for the various cases considered. At PZOA approximation,
BH\&HLYP predicts slightly greater corrections (in magnitude) than DHF, which is explained by the fact that the BH\&HLYP energy gap $\epsilon_l - \epsilon_{h-1}$ that enters  the denominator of eq.\eqref{eq:HL-amp-aprox-QED-PZOA} is, on average, 32\% smaller than that for DHF; in addition, the constant $\gamma$ appearing in the same
equation does not differ significantly between BH\&HLYP and DHF. The opposite is observed with RPA results, since the DHF factors $\mathbb{R}$ appearing in eq. \eqref{eq:HL-amp-aprox-QED-RPA} are 2.1 times greater, on average, than their BH\&HLYP counterparts. Thus, the difference between both approximations in the calculation of
$\delta^\mathrm{QED}(^1 K_\mathrm{iso})$ is twice as large for DHF than for BH\&HLYP wave function.

\begin{figure}[h!]
\centering
\includegraphics[scale=0.7]{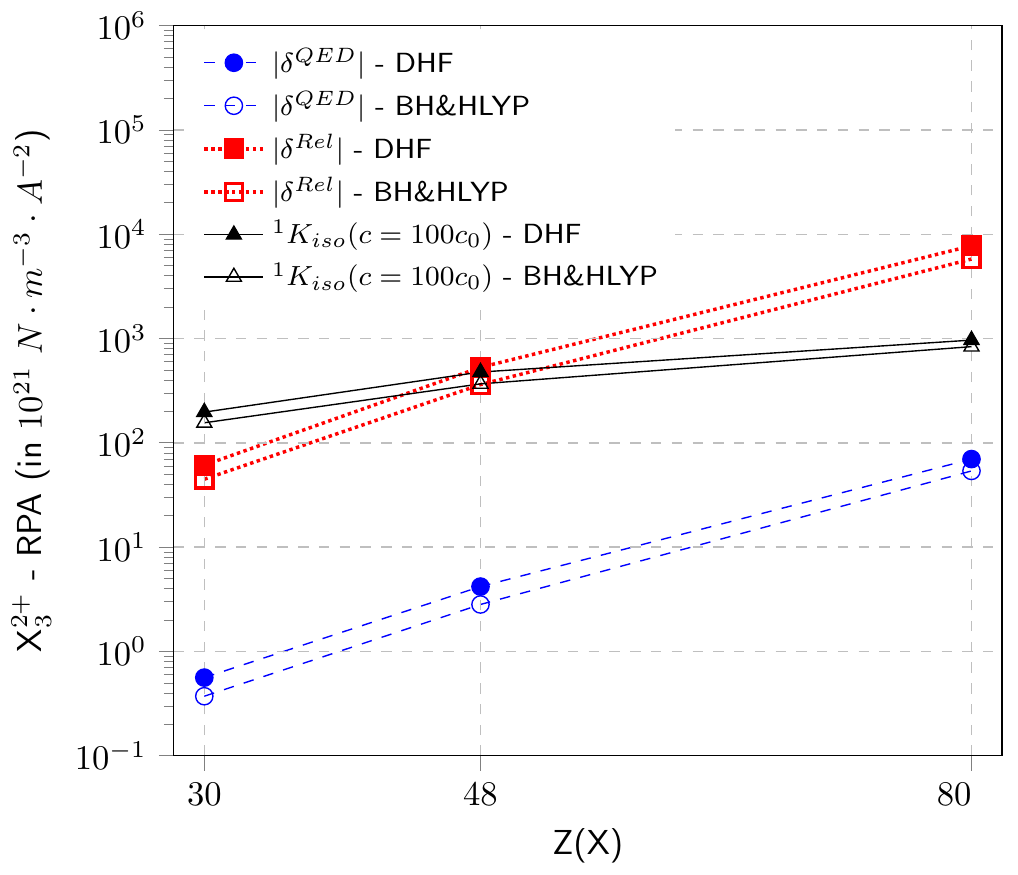}
\caption{Absolute value of QED corrections $|\delta^\mathrm{QED}(^1 K_\mathrm{iso})|$ (in blue) and relativistic effects $\delta^\mathrm{Rel}(^1 K_\mathrm{iso})$ (in red) on the isotropic one-bond coupling, together with the non-relativistic contribution ${^1 K_\mathrm{iso}}(c = 100c_0)$ (in black) in X$_3^{2+}$ ions, plotted as a function of the atomic number Z(X). RPA-DHF values are identified with filled markers and RPA-BH\&HLYP ones with hollow markers. All values are presented in units of $10^{21} N \cdot m^{-3} \cdot A^{-2}$.}
\label{fig:X3-I-QED}
\end{figure}

In Fig. \ref{fig:X3-I-QED} we show RPA-DHF and RPA-BH\&HLYP values for: (absolute value of) QED corrections $|\delta^\mathrm{QED}(^1 K_\mathrm{iso})|$ and relativistic effects $\delta^\mathrm{Rel}(^1 K_\mathrm{iso})$ on the isotropic one-bond coupling, together with the non-relativistic contribution ${^1 K_\mathrm{iso}}(c = 100c_0)$ in X$_3^{2+}$ ions, as a function of the atomic number Z(X). It can be seen that QED corrections remain two orders of magnitude below relativistic effects in the whole range of Z(X). This difference shortens when the comparison is made with the non-relativistic contribution, such that for Hg$_3^{2+}$ the QED correction term is only one order of magnitude smaller than the NR contribution.
%
\subsubsection{One-bond couplings in X$_2^{2+}$}
%
The results for X$_2^{2+}$ ions closely resemble those corresponding to their X$_3^{2+}$ counterparts and a similar analysis can be made following that of the preceding section. Table \ref{tab:QED-eff} reveals that PZOA-BH\&HLYP values exceeds PZOA-DHF ones since the BH\&HLYP energy gap $\epsilon_l - \epsilon_h$ is 36\% lower with respect to that corresponding to DHF. This is reverted at RPA level, since the DHF $\mathbb{R}$ factors are nearly 4 times greater than their BH\&HLYP counterparts.

A plot, similar to Fig. \ref{fig:X3-I-QED}, is given in the Supplementary Information for the X$_2^{2+}$ case. When compared with relativistic effects, QED corrections are two orders of magnitude below in the whole range of the atomic number Z(X). Moreover, the difference between QED corrections and non-relativistic contributions gets even smaller in this case at Hg$_2^{2+}$, being less than one order of magnitude. The fact that RPA QED corrections in X$_2^{2+}$ are one order of magnitude greater than in X$_3^{2+}$ ions is particularly noted.

%
\subsection{QED corrections relative to relativistic effects and total spin-spin couplings}\label{sec:QED-rel-tot}
\begin{figure}[h!]
\centering
\includegraphics[scale=0.7]{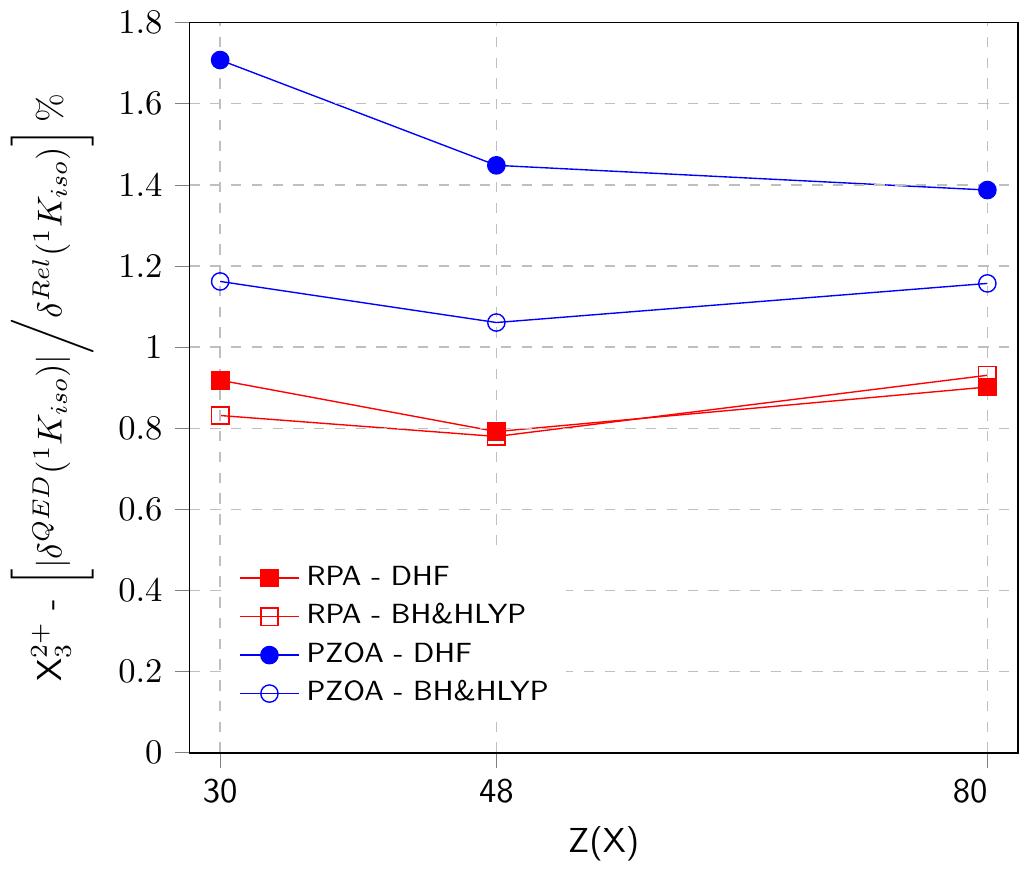}
\caption{Percentage of QED correction $|\delta^\mathrm{QED}(^1 K_\mathrm{iso})|$ relative to total relativistic effect $\delta^\mathrm{Rel}(^1 K_\mathrm{iso})$ in X$_3^{2+}$ ions, as a function of the atomic number of element X. PZOA values are plotted in blue and RPA ones, in red. Filled markers correspond to DHF values, whereas hollow markers correspond to BH\&HLYP ones.}
\label{fig:QED-Rel-X3-I}
\end{figure}

%
In Fig. \ref{fig:QED-Rel-X3-I} we plot the percentage of QED
corrections (in absolute value), $|\delta^\mathrm{QED}(^1 K_\mathrm{iso})|$, to
relativistic effects, $\delta^\mathrm{Rel}(^1 K_\mathrm{iso})$, (for DHF and
BH\&HLYP wave functions and PZOA and RPA approximations) in X$_3^{2+}$
ions, as a function of the atomic number of element X. The use of DHF
or BH\&HLYP wave functions introduces a sizeable difference in the
ratio when PZOA approximation is employed, but this effect is notably
reduced at RPA level. In the latter case, the ratio exhibits a small
variation around 0.8\% within the whole Z range, and approaches 1\% at
$Z = 80$, where good accordance between RPA-DHF and RPA-BH\&HLYP is
achieved. The plot reveals a seeming proportionality relation between
calculated QED corrections and relativistic effects in one-bond
coupling $^1 K_\mathrm{iso}$ at RPA level, a statement that becomes clearer
from Fig. \ref{fig:X3-I-QED} (and its X$_2^{2+}$ counterpart), where
the aforementioned quantities display a functional variation according
to $Z^5$ (see Table \ref{tab:regression}).

More insight on the described functional form for $|\delta^\mathrm{QED}(^1
K_\mathrm{iso})|$ can be given by eq. \eqref{eq:HL-amp-aprox-QED-PZOA}.
Hyperfine integrals $\langle V \rangle_{ns}$ and scaling factors
$\nu_{ns}^\mathrm{QED}$ turn out to be the two main contributions in the
calculation of $|\delta^\mathrm{QED}|$. As noted above, $\langle V \rangle_{ns} \propto Z^2$; additionally, it is found that
$\nu_{ns}^\mathrm{QED} \approx nZ$ (see table \ref{tab:regression}), where the small proportionality constant
($n \sim 10^{-5}$) ensures that only the term $2\nu_{ns}^\mathrm{QED}$
between brackets in eq. \eqref{eq:HL-amp-aprox-QED-PZOA} gives the
major contribution. This ensure that the ratio $\gamma \left( \langle V \rangle_{ns}\right)^2 \nu_{ns}^\mathrm{QED}/(\epsilon_l - \epsilon_h)$ varies according to $Z^5$.

\begin{figure}[h!]
\centering
\includegraphics[scale=0.7]{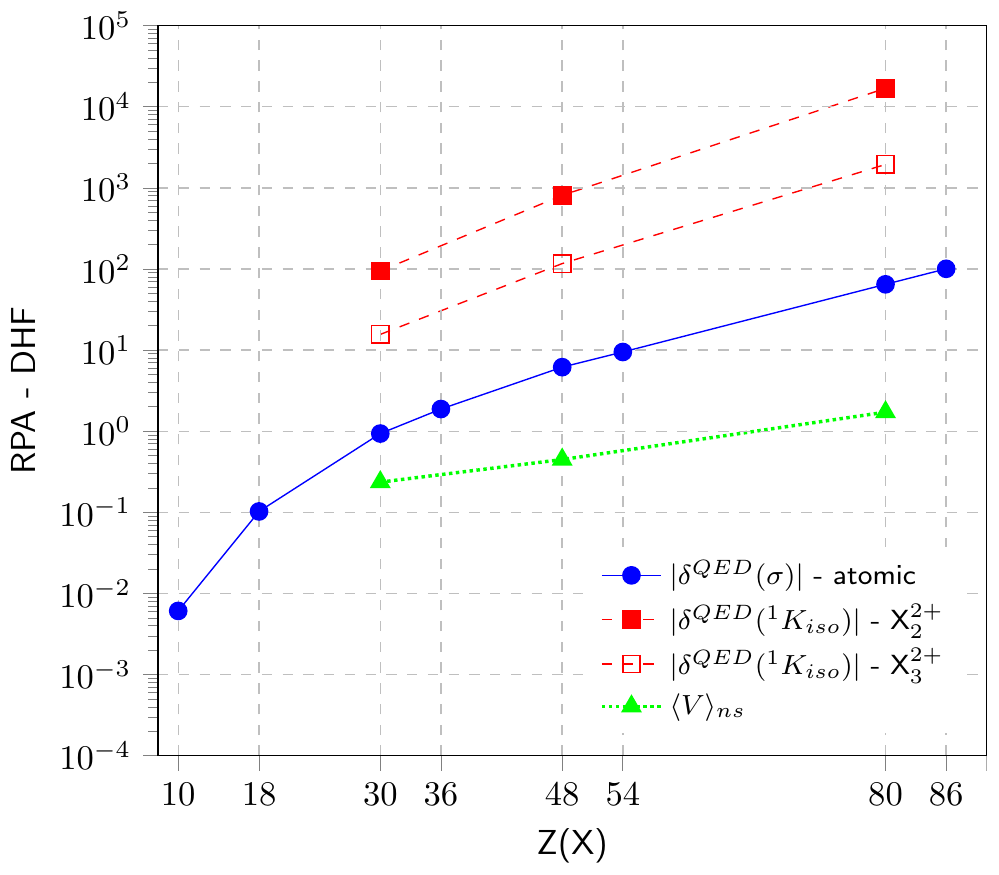}
\caption{Comparison of RPA-DHF values of calculated QED corrections on one-bond coupling $|\delta^\mathrm{QED}(^1 K_\mathrm{iso})|$ (in atomic units $a_0^{-2} \left(\hbar/E_h\right)^{-2} E_h^{-1}$) in X$_2^{2+}$ and X$_3^{2+}$ ions (in red) with the analogous QED corrections on atomic shieldings (in ppm), extracted from Ref. \citenum{Karol_JCP2019} (in blue) as a function of the atomic number Z of the elements involved. We also show the Z-dependence of the hyperfine integrals $\langle V \rangle_{ns}$ (in green and in atomic units $a_0^{-2}$).}
\label{fig:QED-J-shield}
\end{figure}

\begin{figure}[h!]
\centering
\includegraphics[scale=0.7]{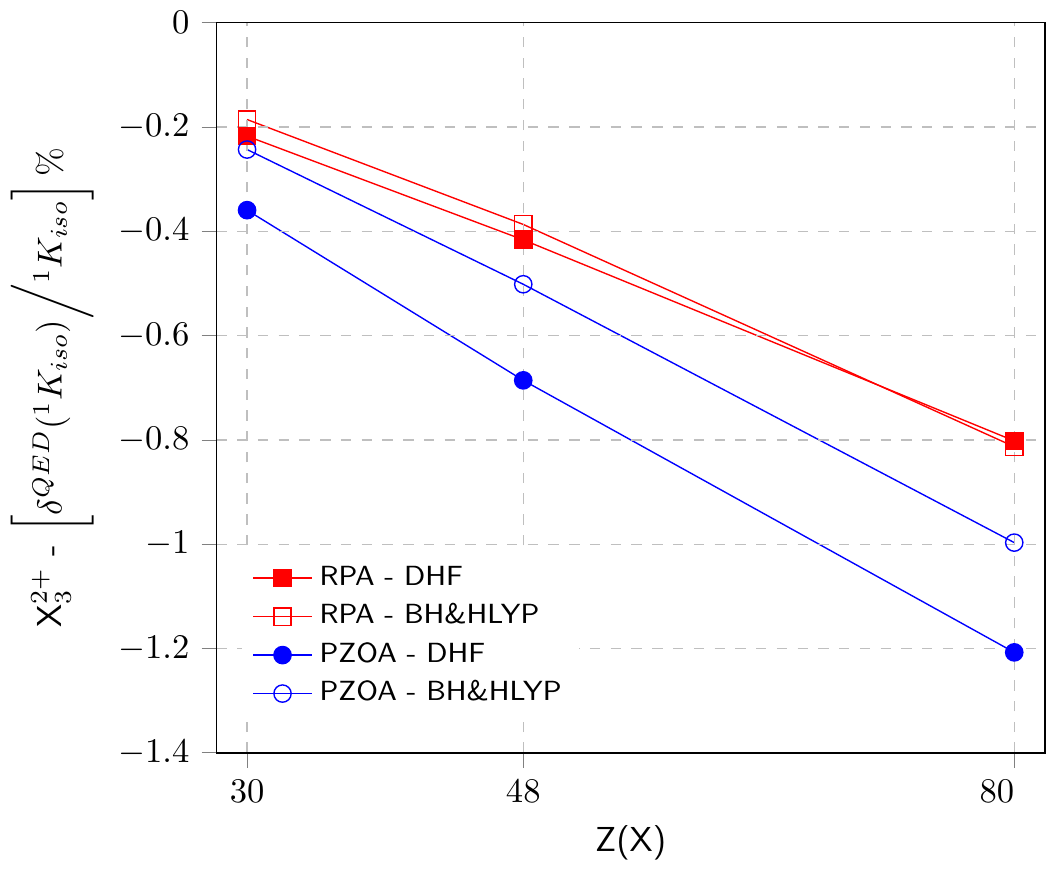}
\caption{Percent contribution of QED corrections to total isotropic one-bond coupling $|\delta^\mathrm{QED}(^1 K_\mathrm{iso})|/{^1 K_\mathrm{iso}}$ in X$_3^{2+}$ ions, as a function of the atomic number of element X. PZOA values are plotted in blue and RPA ones, in red. Filled markers correspond to DHF values and hollow markers correspond to BH\&HLYP.}
\label{fig:QED-Tot-RPA}
\end{figure}

The $Z^5$ dependence of QED corrections $|\delta^\mathrm{QED}(^1 K_\mathrm{iso})|$ closely resembles the results obtained for atomic NMR shieldings, studied in a previous work.\cite{Karol_JCP2019} There, it was shown that estimated QED effects on shielding, affecting matrix elements of both principal propagator and perturbators, vary according to a power law of $Z^5$, with $Z$ the atomic number of the neutral atom. In addition, the major contributions to the response function in the case of electronic shielding were shown to come from interactions between inner core and highly excited atomic orbitals, whereas in the present case, the main contributions to spin-spin coupling arise from valence molecular orbitals, such as HOMO and LUMO. In both cases, $s$-type atomic orbitals proved to have a predominant role ($1s$ and $2s$ in the case of shieldings, $4s$, $5s$ and $6s$ in the present studied ions) and may be behind the observed trend in the calculated QED corrections. In Fig. \ref{fig:QED-J-shield}, QED corrections on one-bond coupling $|\delta^\mathrm{QED}(^1 K_\mathrm{iso})|$ (for both X$_2^{2+}$ and X$_3^{2+}$ ions) are presented together with the analogous corrections on atomic shieldings (extracted from Ref. \citenum{Karol_JCP2019}). The Z-dependence of hyperfine integrals $\langle V \rangle_{ns} \propto Z^2$ is also shown.
Finally, in Fig. \ref{fig:QED-Tot-RPA} we show the (percent) contribution of QED corrections to total isotropic one-bond coupling $|\delta^\mathrm{QED}(^1 K_\mathrm{iso})|/{^1 K_\mathrm{iso}}$ in X$_3^{2+}$ ions. Values for both, DHF and BH\&HLYP, at PZOA and RPA approximations are compared as a function of Z(X). As in Fig. \ref{fig:QED-Rel-X3-I}, a coincidence between DHF and BH\&HLYP when RPA approximation is employed is found in this case. In general, however, an increase of the relative contribution of QED corrections for heavier constituents is described in all cases, reaching $\sim -0.8\%$ for RPA results in mercury ions. This trend is similar to that found for atomic shieldings.\cite{Karol_JCP2019} An analogous plot for X$_2^{2+}$ ions can be found in the Supplementary Information.
\begin{table*}[ht!]
\setlength{\extrarowheight}{1.5mm}
\caption[]{Regression parameters for: total isotropic one-bond indirect nuclear spin-spin coupling $^1 K_\mathrm{iso}$, both for $c = c_0$ and $c = 100c_0$, QED corrections to total isotropic one-bond coupling (in absolute value), $|\delta^\mathrm{QED}(^1 K_\mathrm{iso})|$, relativistic effect on total isotropic one-bond coupling, $\delta^\mathrm{Rel}(^1 K_\mathrm{iso})$ and atomic hyperfine integrals, $\langle V \rangle_{ns}$, as a function of the atomic number of element X. For these quantities, a power law of the form $Q = mZ^n$ is proposed (or its equivalent linearized form $\ln Q = \ln m + n\ln Z$). Items indicate the set of values used for the regression: a) PZOA - DHF; b) PZOA - BH\&HLYP; c) RPA - DHF; d) RPA - BH\&HLYP. In the last row, linear regression parameters for the relation $\nu^\mathrm{QED} = m + nZ$ ($\nu^\mathrm{QED}$ from Ref. \citenum{Sapirstein_PRA2003}) as a function of $Z$ (these last results show an almost proportional relation $\nu^\mathrm{QED} \propto Z$).}
\label{tab:regression}
\begin{tabular*}{\linewidth}{@{\extracolsep{\fill}} *{6}{c}}
\hline\hline
 & \multicolumn{5}{c}{$\ensuremath{\ln Q = \ln m + n\ln Z}$} \\
\cline{2-6}
$\ensuremath{Q}$ &  &  & $\ensuremath{\ln m}$ & $\ensuremath{n}$ & $\ensuremath{R^2}$ \\
\hline
\multirow{8}{3cm}{$\ensuremath{^1 K_\mathrm{iso}(c = c_0)}$} & \multirow{4}{0.3cm}{X$\ensuremath{_2^{2+}}$} & a) & $\ensuremath{-10 \pm 2}$ & $\ensuremath{4.0 \pm 0.6}$ & $\ensuremath{0.9756}$ \\
 &  & b) & $\ensuremath{-8 \pm 2}$ & $\ensuremath{3.8 \pm 0.6}$ & $\ensuremath{0.9775}$ \\
 &  & c) & $\ensuremath{-7 \pm 2}$ & $\ensuremath{4.0 \pm 0.4}$ & $\ensuremath{0.9899}$ \\
 &  & d) & $\ensuremath{-6 \pm 2}$ & $\ensuremath{3.8 \pm 0.5}$ & $\ensuremath{0.9835}$ \\
\cline{2-6}
 & \multirow{4}{0.3cm}{X$\ensuremath{_3^{2+}}$} & a) & $\ensuremath{-10 \pm 2}$ & $\ensuremath{3.9 \pm 0.6}$ & $\ensuremath{0.9783}$ \\
 &  & b) & $\ensuremath{-9 \pm 2}$ & $\ensuremath{3.7 \pm 0.5}$ & $\ensuremath{0.9797}$ \\
 &  & c) & $\ensuremath{-7 \pm 2}$ & $\ensuremath{3.6 \pm 0.4}$ & $\ensuremath{0.9888}$ \\
 &  & d) & $\ensuremath{-7 \pm 2}$ & $\ensuremath{3.6 \pm 0.5}$ & $\ensuremath{0.9843}$ \\
\hline
\multirow{8}{3cm}{$\ensuremath{^1 K_\mathrm{iso}(c = 100c_0)}$} & \multirow{4}{0.3cm}{X$\ensuremath{_2^{2+}}$} & a) & $\ensuremath{-3.2 \pm 0.3}$ & $\ensuremath{2.07 \pm 0.07}$ & $\ensuremath{0.9988}$ \\
 &  & b) & $\ensuremath{-1.9 \pm 0.3}$ & $\ensuremath{1.95 \pm 0.06}$ & $\ensuremath{0.9989}$ \\
 &  & c) & $\ensuremath{2.3 \pm 0.8}$ & $\ensuremath{1.3 \pm 0.2}$ & $\ensuremath{0.9777}$ \\
 &  & d) & $\ensuremath{0.9 \pm 0.3}$ & $\ensuremath{1.61 \pm 0.07}$ & $\ensuremath{0.9981}$ \\
\cline{2-6}
 & \multirow{4}{0.3cm}{X$\ensuremath{_3^{2+}}$} & a) & $\ensuremath{-4 \pm 0.2}$ & $\ensuremath{2.06 \pm 0.04}$ & $\ensuremath{0.9996}$ \\
 &  & b) & $\ensuremath{-2.7 \pm 0.1}$ & $\ensuremath{1.93 \pm 0.03}$ & $\ensuremath{0.9997}$ \\
 &  & c) & $\ensuremath{-0.2 \pm 0.6}$ & $\ensuremath{1.6 \pm 0.1}$ & $\ensuremath{0.9921}$ \\
 &  & d) & $\ensuremath{-0.8 \pm 0.2}$ & $\ensuremath{1.71 \pm 0.06}$ & $\ensuremath{0.9986}$ \\
\hline
\multirow{8}{3cm}{$\ensuremath{|\delta^\mathrm{QED}(^1 K_\mathrm{iso})|}$} & \multirow{4}{0.3cm}{X$\ensuremath{_2^{2+}}$} & a) & $\ensuremath{-19 \pm 2}$ & $\ensuremath{5.2 \pm 0.6}$ & $\ensuremath{0.9872}$ \\
 &  & b) & $\ensuremath{-15.97}$ & $\ensuremath{4.35}$ & $\ensuremath{1}$ \\
 &  & c) & $\ensuremath{-17 \pm 2}$ & $\ensuremath{5.3 \pm 0.4}$ & $\ensuremath{0.9944}$ \\
 &  & d) & $\ensuremath{-14.97}$ & $\ensuremath{4.51}$ & $\ensuremath{1}$ \\
\cline{2-6}
 & \multirow{4}{0.3cm}{X$\ensuremath{_3^{2+}}$} & a) & $\ensuremath{-20 \pm 2}$ & $\ensuremath{5.1 \pm 0.5}$ & $\ensuremath{0.9904}$ \\
 &  & b) & $\ensuremath{-20 \pm 2}$ & $\ensuremath{5.2 \pm 0.5}$ & $\ensuremath{0.9914}$ \\
 &  & c) & $\ensuremath{-17 \pm 1}$ & $\ensuremath{4.9 \pm 0.4}$ & $\ensuremath{0.9948}$ \\
 &  & d) & $\ensuremath{-18 \pm 2}$ & $\ensuremath{5.1 \pm 0.4}$ & $\ensuremath{0.9933}$ \\
\hline
\multirow{8}{3cm}{$\ensuremath{\delta^\mathrm{Rel}(^1 K_\mathrm{iso})}$} & \multirow{4}{0.3cm}{X$\ensuremath{_2^{2+}}$} & a) & $\ensuremath{-16 \pm 2}$ & $\ensuremath{5.4 \pm 0.5}$ & $\ensuremath{0.9925}$ \\
 &  & b) & $\ensuremath{-14 \pm 2}$ & $\ensuremath{5.2 \pm 0.4}$ & $\ensuremath{0.9941}$ \\
 &  & c) & $\ensuremath{-11.1 \pm 0.7}$ & $\ensuremath{5.1 \pm 0.2}$ & $\ensuremath{0.9989}$ \\
 &  & d) & $\ensuremath{-12 \pm 1}$ & $\ensuremath{5.0 \pm 0.3}$ & $\ensuremath{0.9968}$ \\
\cline{2-6}
 & \multirow{4}{0.3cm}{X$\ensuremath{_3^{2+}}$} & a) & $\ensuremath{-17 \pm 2}$ & $\ensuremath{5.4 \pm 0.4}$ & $\ensuremath{0.9936}$ \\
 &  & b) & $\ensuremath{-15 \pm 1}$ & $\ensuremath{5.1 \pm 0.4}$ & $\ensuremath{0.9947}$ \\
 &  & c) & $\ensuremath{-12.7 \pm 0.7}$ & $\ensuremath{4.9 \pm 0.2}$ & $\ensuremath{0.9985}$ \\
 &  & d) & $\ensuremath{-13 \pm 1}$ & $\ensuremath{5.0 \pm 0.3}$ & $\ensuremath{0.9968}$ \\
\hline
$\langle V \rangle_{ns}$ &  &  & $\ensuremath{-8 \pm 1}$ & $\ensuremath{2.0 \pm 0.4}$ & 0.9689 \\
\hline
\hline
&  &  & $m$ & $n$ & $R^2$ \\
\hline
$\nu^\mathrm{QED} = m + nZ$ &  &  & $(8 \pm 2) 10^{-4}$ & $(-8.3 \pm 0.5) 10^{-5}$ & $0.9834$ \\
\hline
\hline
\end{tabular*}
\end{table*}
%
\section{Conclusions}
%
When searching for highly accurate atomic and molecular response properties one should nowadays consider physical effects that were taken to be vanishingly small few years ago. Among them, one must include QED effects and Breit interactions. 
In the present work, we proposed an effective model that can be applied to simple molecular systems in order to estimate QED corrections to the indirect nuclear spin-spin coupling. This model permitted an estimation of the order of magnitude in the case of one-bond indirect nuclear spin-spin couplings in X$_2^{2+}$ and X$_3^{2+}$ ions (X = Zn, Cd, Hg), within polarization propagator theory, using RPA and PZOA approximations and DHF and BH\&HLYP wavefunctions, both at four-component and ZORA levels.
The similitude between orders of magnitude obtained with different levels of theory, confirms the consistency of our approach to estimate the QED effect on the NMR indirect nuclear spin-spin coupling constant. The QED corrections were found in the interval $(0.7; ~ 1.7)$\% of the total relativistic effect on $^1 K_\mathrm{iso}$ in X$_2^{2+}$ and X$_3^{2+}$ ions and from the interval $(-0.2; ~ -0.4)$\% in Zn-containing ions to $(-0.8; ~ -1.2)$\% in Hg-containing ions (with visible $Z$-dependence) of the total isotropic one-bond coupling constant.

At the moment, we are trying to extend the application of our models
to  more complex molecules, with the dominant $p$-type atomic orbital component in HOMO/LUMO molecular orbitals. 
Our results show that QED effects can be sizeable in heavy atom-containing compounds and its consideration may enhance the theoretical prediction of measured values in highly accurate experiments, since the QED correction to indirect nuclear spin-spin coupling is in the order of the experimental uncertainty of $J$ for high-$Z$ atom-containing molecules.

As a further step, we aim to the inclusion of solvent effects in our calculations. This will allow us to compare our theoretical predictions with measured values obtained in liquid-phase experiments. We hope that the present work will inspire  future experiments with higher accuracy. Zero and ultra-low field (ZULF) NMR measurements\cite{ledbetter:2011} that can achieve resolution of mHz on indirect nuclear spin-spin couplings\cite{wilzewski:2017} are particularly promising for this purpose if they were performed on heavy-element containing systems.


\section*{Acknowledgements}
We acknowledge support from CONICET by grant PIP 112-20130100361 and FONCYT by grant PICT 2016-2936 as well as funding by the Deutsche Forschungsgemeinschaft (DFG, German Research Foundation) -- Projektnummer 445296313. Computer time provided by the center for scientific computing (CSC) Frankfurt is gratefully acknowledged.

%
%

\bibliographystyle{aipnum4-1}

\bibliography{QED_J}

%
%


\end{document}


\title{Relativistic and QED corrections to one-bond indirect
nulcear spin-spin couplings in X$_2^{2+}$ and X$_3^{2+}$ ions (X = Zn, Cd, Hg) \\ Supplementary Information}

\author{Mariano Colombo Jofr\'e}%
\affiliation{Instituto de Modelado e Innovaci\'on Tecnol\'ogica (IMIT), Facultad de Ciencias Exactas, Naturales y Agrimensura, Universidad Nacional del Nordeste, Avda. Libertad 5460, W3404AAS, Corrientes, Argentina}

\author{Karol Kozio{\l}}
\affiliation{Narodowe Centrum Bada\'{n} J\k{a}drowych (NCBJ), Andrzeja So{\l}tana 7, 05-400 Otwock-\'{S}wierk, Poland}

\author{I. Agust{\'i}n Aucar}
\affiliation{Instituto de Modelado e Innovaci\'on Tecnol\'ogica (IMIT), Facultad de Ciencias Exactas, Naturales y Agrimensura, Universidad Nacional del Nordeste, Avda. Libertad 5460, W3404AAS, Corrientes, Argentina}

\author{Konstantin Gaul}
\author{Robert Berger}
\affiliation{Fachbereich Chemie, Philipps–Universit\"at Marburg, Hans-Meerwein-Stra{\ss}e 4, 35032 Marburg, Germany}

\author{Gustavo A. Aucar}
\email{gaa@unne.edu.ar}
\affiliation{Instituto de Modelado e Innovaci\'on Tecnol\'ogica (IMIT), Facultad de Ciencias Exactas, Naturales y Agrimensura, Universidad Nacional del Nordeste, Avda. Libertad 5460, W3404AAS, Corrientes, Argentina}


\date{\today}

\keywords{Relativistic effects, ee and pp contributions, diatomic molecules, triatomic molecules, QED}
\maketitle



\begin{table}[h!]
\caption{Optimized internuclear distances (in \AA) in the various ions treated in this work. All values were obtained using the Dirac-Coulomb Hamiltonian and the unrestricted kinetic balance. dyall.cv3z basis set was used in all cases.}
\setlength{\extrarowheight}{1.5mm}
\centering
\begin{tabular*}{0.6\linewidth}{@{\extracolsep{\fill}}c *{2}{c}}
X & X$_2^{2+}$ & X$_3^{2+}$ \\
\hline
\hline
Zn & 2.5253 & 2.5587 \\
Cd & 2.8253 & 2.8499 \\
Hg & 2.6896 & 2.7124 \\
\hline
\hline
\end{tabular*}
\end{table}

\begin{table}[h!]
\caption{Numerical values of radial integrals used in this work, obtained from MCDFGME code (in atomic units $a_0^{-2}$).}
\setlength{\extrarowheight}{1.5mm}
\centering
\begin{tabular*}{0.3\linewidth}{@{\extracolsep{\fill}}c S[round-mode=places,round-precision=4]}
X & {$\langle V \rangle_{ns}$} \\
\hline
\hline
Zn & 0,23572553 \\
Cd & 0,44785723 \\
Hg & 1,7166458 \\
\hline
\hline
\end{tabular*}
\end{table}

\begin{figure}[ht!]
\centering
\includegraphics[scale=0.7]{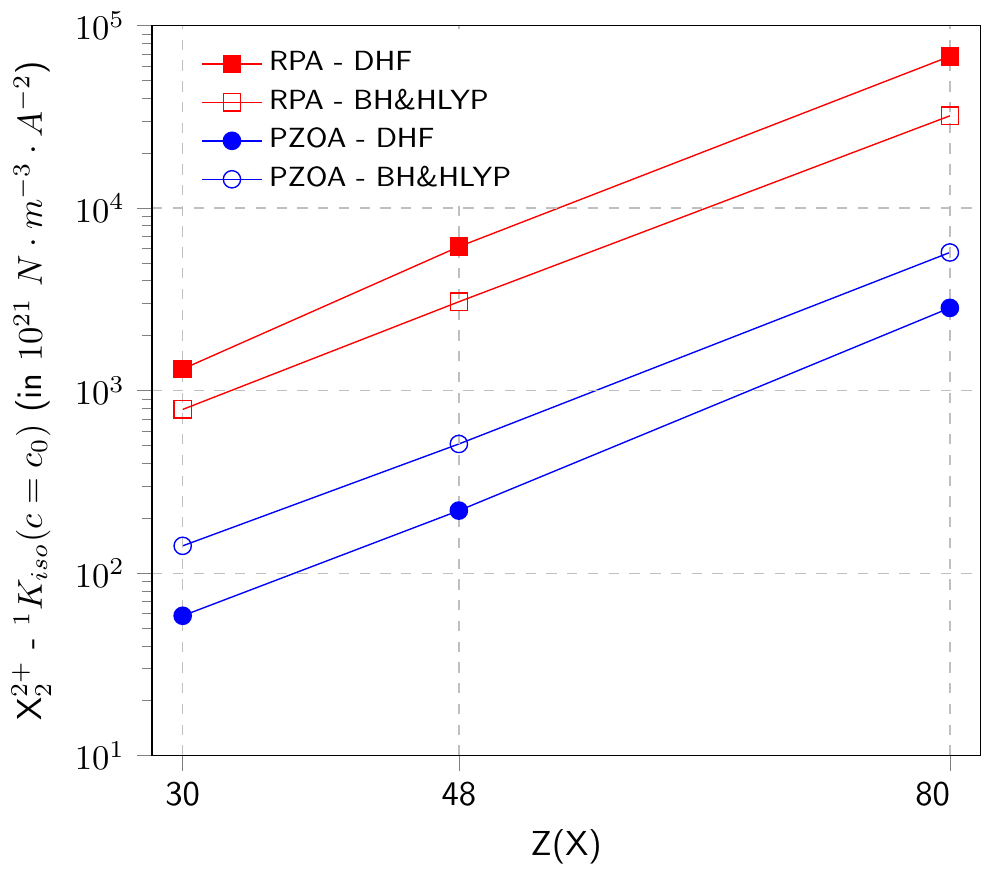}
\caption{Total values of one-bond isotropic reduced coupling $^1 K_{iso}$ (in $10^{21} ~ \mathrm{N} \cdot \mathrm{m}^{-3} \cdot \mathrm{A}^{-2}$) in X$_2^{2+}$ ions, as a function of the atomic number of element X, Z(X). PZOA values are plotted in blue and RPA ones, in red. Filled markers correspond to DHF results and hollow markers correspond to BH\&HLYP ones. A logarithmic scale was set for the vertical axis, thus indicating a power-law as a function of Z(X).}
\label{}
\end{figure}

\begin{figure}[ht!]
\centering
\includegraphics[scale=0.7]{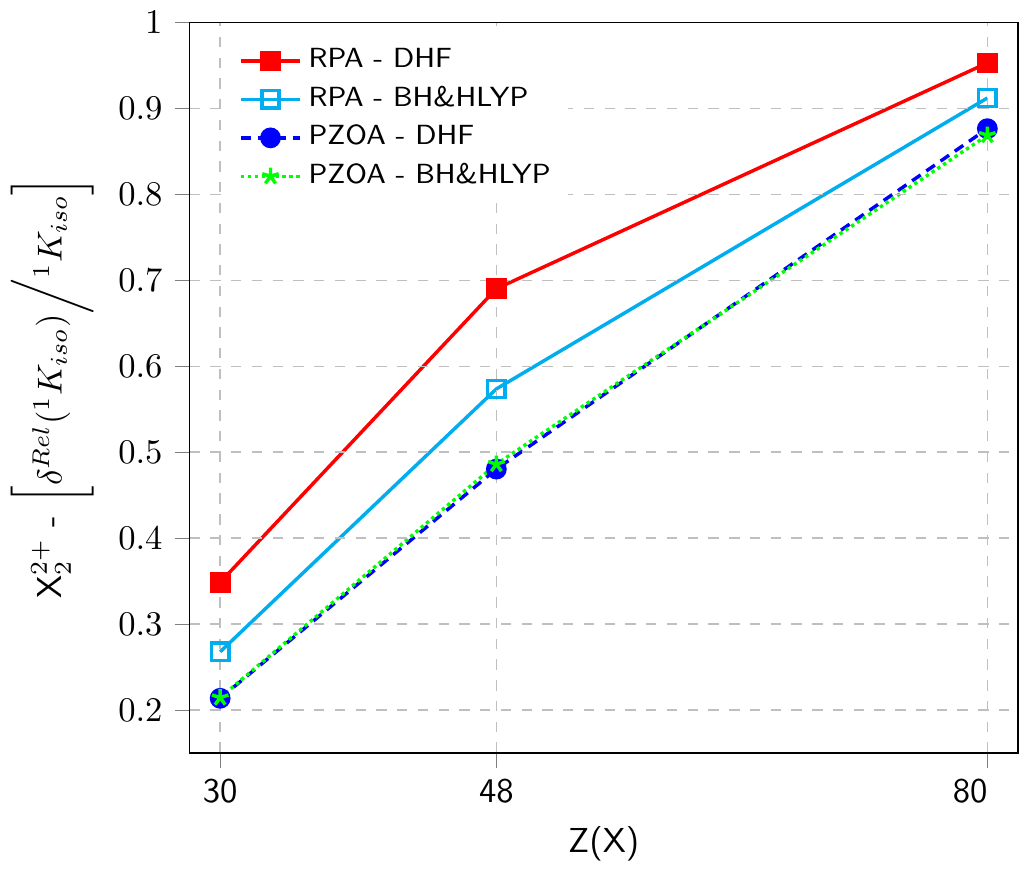}
\caption{Ratio of the relativistic effect to the total value of isotropic one-bond reduced coupling $\delta^{Rel}(^1 K_{iso})/{^1 K_{iso}}$ as a function of the atomic number Z(X) in X$_2^{2+}$ ions. PZOA-DHF values are plotted with blue dashed line and filled circles and PZOA-BH\&HLYP with green dotted line and stars (the curves overlap in part). In red solid line with filled squares, the values for RPA-DHF and the cyan solid line with hollow squares, the RPA-BH\&HLYP ones.}
\label{X3-II-PZOA}
\end{figure}

\begin{figure}[ht!]
\centering
\includegraphics[scale=0.7]{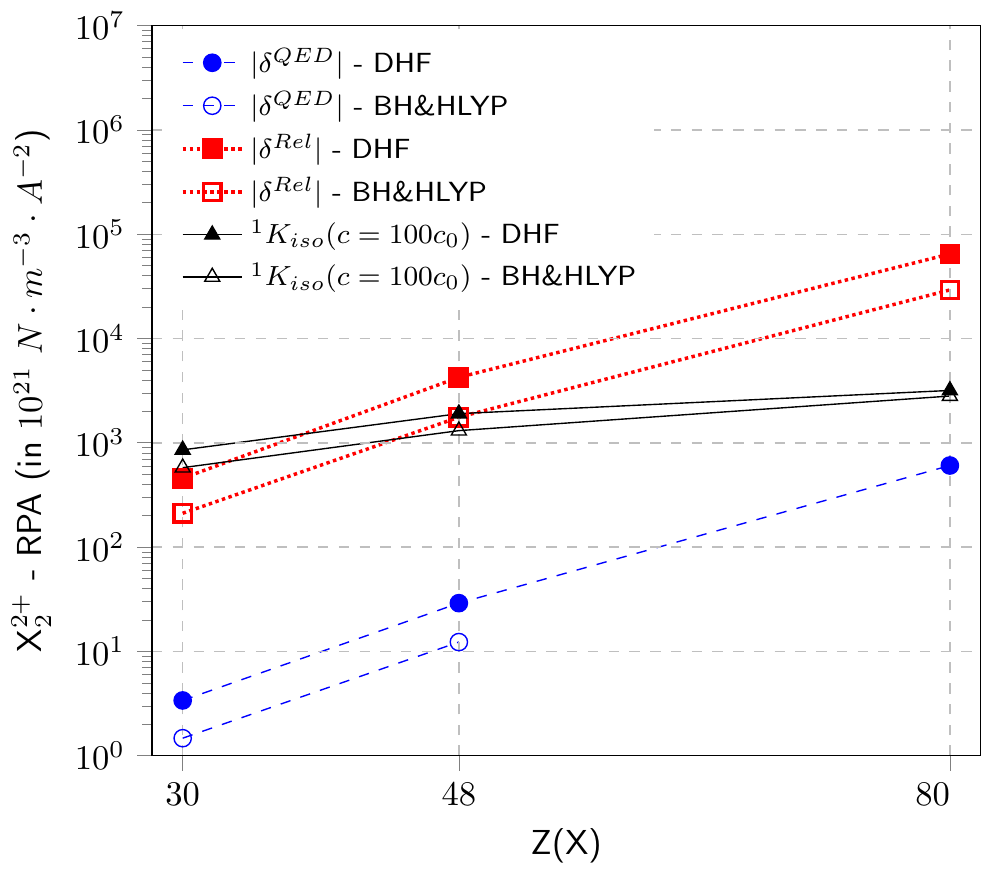}
\caption{Absolute value of QED corrections $|\delta^{QED}(^1 K_{iso})|$ (in blue) and relativistic effects $\delta^{Rel}(^1 K_{iso})$ (in red) on the isotropic one-bond coupling, together with the non-relativistic contribution ${^1 K_{iso}}(c = 100c_0)$ (in black) in X$_2^{2+}$ ions, plotted as a function of the atomic number Z(X). RPA-DHF values are identified with filled markers and RPA-BH\&HLYP ones with hollow markers. All values are presented in units of $10^{21} ~ N \cdot m^{-3} \cdot A^{-2}$.}
\label{X3-II-PZOA}
\end{figure}

\begin{figure}[ht!]
\centering
\includegraphics[scale=0.7]{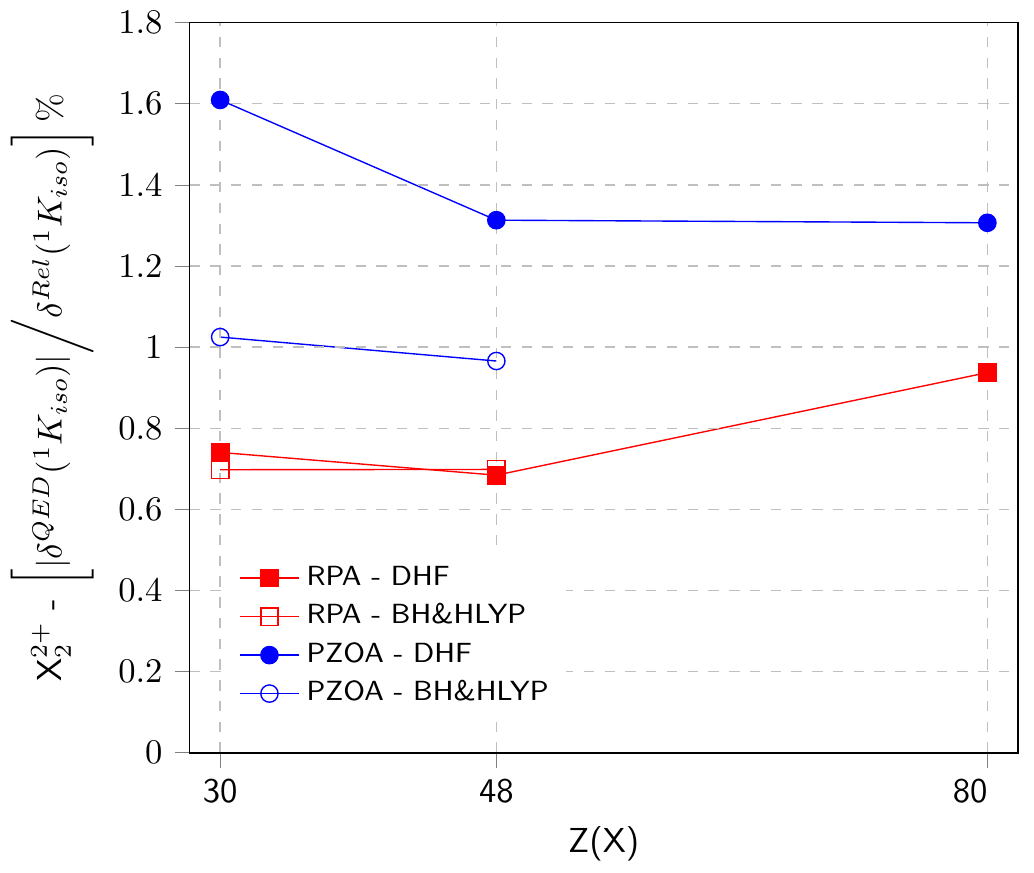}
\caption{Percentage of QED correction $|\delta^{QED}(^1 K_{iso})|$ relative to total relativistic effect $\delta^{Rel}(^1 K_{iso})$ in X$_2^{2+}$ ions, as a function of the atomic number of element X. PZOA values are plotted in blue and RPA ones, in red. Filled markers correspond to DHF results and hollow markers correspond to BH\&HLYP ones.}
\label{fig:X3-I-Rel}
\end{figure}

\begin{figure}[ht!]
\centering
\includegraphics[scale=0.7]{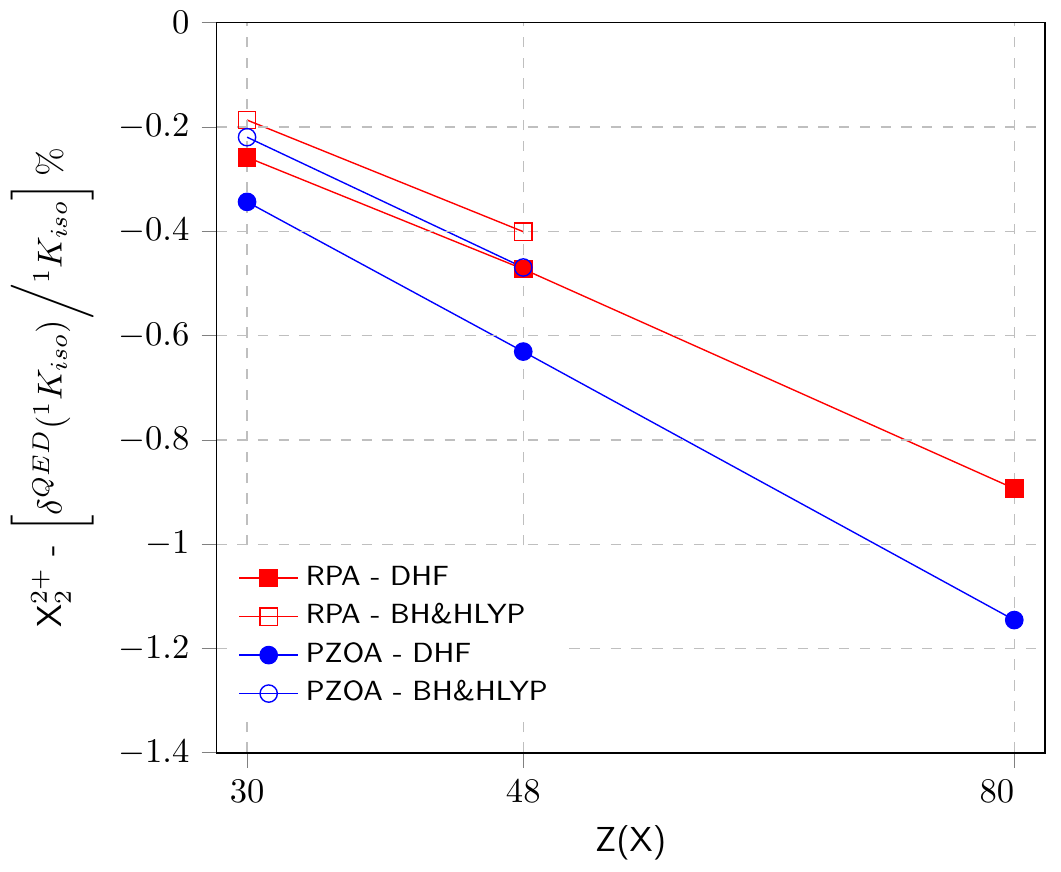}
\caption{Percent contribution of QED corrections to total isotropic one-bond coupling $|\delta^{QED}(^1 K_{iso})|/{^1 K_{iso}}$ in X$_2^{2+}$ ions, as a function of the atomic number of element X. PZOA values are plotted in blue and RPA ones, in red. Filled markers correspond to DHF results and hollow markers correspond to BH\&HLYP ones.}
\label{fig:X3-I-Tot}
\end{figure}

\begin{table*}[h!]
\caption{Projection analysis expansion coefficients for HOMO and LUMO in X$_2^{2+}$ ions, using DHF wave function. $ns$ stands for the valence $s$-type atomic orbital of element X, used as the projection reference orbital. $c$ coefficients corresponds to HOMO. $d$ coefficients corresponds to LUMO.}
\setlength{\extrarowheight}{1.1mm}
\begin{tabular*}{\linewidth}{@{\extracolsep{\fill}}l *{4}{c}}
\hline
\hline
 & $c_{ns}^1$ & $c_{\bar{ns}}^1$ & $c_{ns}^2$ & $c_{\bar{ns}}^2$ \\
\hline
Zn$_2^{2+}$ & $0.0586$ & $(0.0707) + i(0.5778)$ & $0.0586$ & $(0.0707) + i(0.5778)$ \\
Cd$_2^{2+}$ & $-0.0129$ & $(-0.1124) + i(0.5834)$ & $-0.0129$ & $(-0.1124) + i(0.5834)$ \\
Hg$_2^{2+}$ & $0.0323$ & $(0.600) + i(0.0530)$ & $0.0323$ & $(0.600) + i(0.0530)$ \\
\hline
 & $d_{ns}^1$ & $d_{\bar{ns}}^1$ & $d_{ns}^2$ & $d_{\bar{ns}}^2$ \\
\hline
Zn$_2^{2+}$ & $(-0.0564) + i(-0.0346)$ & $(-0.0025) + i(-0.8453)$ & $(0.0564) + i(0.0346)$ & $(0.0025) + i(0.8453)$ \\
Cd$_2^{2+}$ & $(0.0255) + i(0.2283)$ & $(0.7787) + i(0.1721)$ & $(-0.0255) + i(-0.2283)$ & $(-0.7787) + i(-0.1721)$ \\
Hg$_2^{2+}$ & $(-0.0698) + i(0.0157)$ & $(-0.8221) + i(-0.0720)$ & $(0.0698) + i(-0.0157)$ & $(0.8221) + i(0.0720)$ \\
\hline
\hline
\end{tabular*}
\end{table*}

\begin{table*}[h!]
\caption{Projection analysis expansion coefficients for ``HOMO-1'' and LUMO in X$_3^{2+}$ ions, using DHF wave function. $ns$ stands for the valence $s$-type atomic orbital of element X, used as the projection reference orbital. $c$ coefficients corresponds to HOMO-1. $d$ coefficients corresponds to LUMO.}
\setlength{\extrarowheight}{1.1mm}
\begin{tabular*}{\linewidth}{@{\extracolsep{\fill}}l *{4}{c}}
\hline
\hline
 & $c_{ns}^1$ & $c_{\bar{ns}}^1$ & $c_{ns}^2$ & $c_{\bar{ns}}^2$ \\
\hline
Zn$_3^{2+}$ & $-0.0295$ & $(-0.0356) + i(-0.2911)$ & $-0.0656$ & $(-0.0792) + i(-0.6475)$ \\
Cd$_3^{2+}$ & $-0.0066$ & $(-0.0579) + i(0.3003)$ & $-0.0143$ & $(-0.1247) + i(0.6470)$ \\
Hg$_3^{2+}$ & $0.0172$ & $(0.3198) + i(0.0283)$ & $0.0355$ & $(0.6583) + i(0.0582)$ \\
\hline
 & $d_{ns}^1$ & $d_{\bar{ns}}^1$ & $d_{ns}^2$ & $d_{\bar{ns}}^2$ \\
\hline
Zn$_3^{2+}$ & $0.0714$ & $(0.0862) + i(0.7049)$ & $-0.0958$ & $(-0.1156) + i(-0.9451)$ \\
Cd$_3^{2+}$ & $-0.0150$ & $(-0.1308) + i(0.6785)$ & $0.0200$ & $(0.1742) + i(-0.9041)$ \\
Hg$_3^{2+}$ & $-0.0366$ & $(-0.6792) + i(-0.0600)$ & $0.0480$ & $(0.8913) + i(0.0788)$ \\
\hline
\hline
\end{tabular*}
\end{table*}

\begin{table*}[h!]
\caption{Projection analysis expansion coefficients for HOMO and LUMO in X$_2^{2+}$ ions, using BH\&HLYP wave function. $ns$ stands for the valence $s$-type atomic orbital of element X, used as the projection reference orbital. $c$ coefficients corresponds to HOMO. $d$ coefficients corresponds to LUMO.}
\setlength{\extrarowheight}{1.1mm}
\begin{tabular*}{\linewidth}{@{\extracolsep{\fill}}l *{4}{c}}
\hline
\hline
 & $c_{ns}^1$ & $c_{\bar{ns}}^1$ & $c_{ns}^2$ & $c_{\bar{ns}}^2$ \\
\hline
Zn$_2^{2+}$ & $-0.1191$ & $(-0.5702) + i(-0.1173)$ & $-0.1191$ & $(-0.5702) + i(-0.1173)$ \\
Cd$_2^{2+}$ & $-0.3556$ & $(0.4385) + i(0.2131)$ & $-0.3556$ & $(0.4385) + i(0.2131)$ \\
Hg$_2^{2+}$ & $0.0326$ & $(0.6045) + i(0.0534)$ & $0.0326$ & $(0.6045) + i(0.0534)$ \\
\hline
 & $d_{ns}^1$ & $d_{\bar{ns}}^1$ & $d_{ns}^2$ & $d_{\bar{ns}}^2$ \\
\hline
Zn$_2^{2+}$ & $(-0.2720) + i(0.0041)$ & $(-0.6947) + i(-0.3976)$ & $(0.2720) + i(-0.0041)$ & $(0.6947) + i(0.3976)$ \\
Cd$_2^{2+}$ & $(-0.4654) + i(-0.0937)$ & $(0.5827) + i(0.3523)$ & $(0.4654) + i(0.0937)$ & $(-0.5827) + i(-0.3523)$ \\
Hg$_2^{2+}$ & $(-0.5673) + i(0.5392)$ & $(0.2654) + i(0.0531)$ & $(0.5673) + i(-0.5392)$ & $(-0.2654) + i(-0.0531)$ \\
\hline
\hline
\end{tabular*}
\end{table*}

\begin{table*}[h!]
\caption{Projection analysis expansion coefficients for ``HOMO-1'' and LUMO in X$_3^{2+}$ ions, using BH\&HLYP wave function. $ns$ stands for the valence $s$-type atomic orbital of element X, used as the projection reference orbital. $c$ coefficients corresponds to HOMO-1. $d$ coefficients corresponds to LUMO.}
\setlength{\extrarowheight}{1.1mm}
\begin{tabular*}{\linewidth}{@{\extracolsep{\fill}}l *{4}{c}}
\hline
\hline
 & $c_{ns}^1$ & $c_{\bar{ns}}^1$ & $c_{ns}^2$ & $c_{\bar{ns}}^2$ \\
\hline
Zn$_3^{2+}$ & $-0.0608$ & $(-0.2908) + i(-0.0598)$ & $-0.1348$ & $(-0.6451) + i(-0.1327)$ \\
Cd$_3^{2+}$ & $0.1849$ & $(-0.2279) + i(-0.1108)$ & $0.3996$ & $(-0.4927) + i(-0.2395)$ \\
Hg$_3^{2+}$ & $0.0173$ & $(0.3219) + i(0.0284)$ & $0.0360$ & $(0.6689) + i(0.0591)$ \\
\hline
 & $d_{ns}^1$ & $d_{\bar{ns}}^1$ & $d_{ns}^2$ & $d_{\bar{ns}}^2$ \\
\hline
Zn$_3^{2+}$ & $0.1410$ & $(0.6749) + i(0.1388)$ & $-0.1829$ & $(-0.8755) + i(-0.1801)$ \\
Cd$_3^{2+}$ & $0.4041$ & $(-0.4983) + i(-0.2422)$ & $-0.5191$ & $(0.6400) + i(0.3111)$ \\
Hg$_3^{2+}$ & $-0.0365$ & $(-0.6776) + i(-0.0599)$ & $0.0462$ & $(0.8576) + i(0.0758)$ \\
\hline
\hline
\end{tabular*}
\end{table*}